\documentclass[12pt,preprint]{aastex}

\begin{document}

\title{Investigating ChaMPlane X-ray sources in the Galactic Bulge with Magellan LDSS2 spectra}

\author{Xavier Koenig,\altaffilmark{1} Jonathan E. Grindlay,\altaffilmark{1} Maureen van den Berg,\altaffilmark{1} Silas Laycock,\altaffilmark{1} Ping Zhao,\altaffilmark{1} JaeSub Hong\altaffilmark{1} and Eric M. Schlegel\altaffilmark{2}}
\altaffiltext{1}{Harvard-Smithsonian Center for Astrophysics, 60 Garden Street, Cambridge, MA 02138; xkoenig@cfa.harvard.edu}
\altaffiltext{2}{Department of Physics and Astronomy, University of Texas at San Antonio, 1 UTSA Circle, San Antonio, TX 78249}

\begin{abstract}
  We have carried out optical and X-ray spectral analyses on a sample
  of 136 candidate optical counterparts of X-ray sources found in five
  Galactic-bulge fields included in our $\it{Chandra}$
  Multi-wavelength Plane Survey. We use a combination of optical
  spectral fitting and quantile X-ray analysis to obtain the hydrogen
  column density towards each object, and a three-dimensional dust
  model of the Galaxy to estimate the most probable distance in each
  case. We present the discovery of a population of stellar coronal
  emission sources, likely consisting of pre-main sequence, young main
  sequence and main sequence stars, as well as a component of active
  binaries of RS CVn or BY Dra type. We identify one candidate
  quiescent low-mass X-ray binary with a sub-giant companion; we note
  that this object may also be an RS CVn system. We report the
  discovery of three new X-ray detected cataclysmic variables (CVs) in
  the direction of the Galactic Center (at distances $\lesssim$
  2kpc). This number is in excess of predictions made with a simple CV
  model based on a local CV space density of $\lesssim$ 10$^{-5}$
  pc$^{-3}$, and a scale height $\sim$200pc. We discuss several
  possible reasons for this observed excess.
\end{abstract}


\section{Introduction}
The goal of the $\it{Chandra}$ Multi-wavelength Plane Survey
(ChaMPlane)\footnote{http://hea-www.harvard.edu/ChaMPlane} is to study
the galactic X-ray point-source population, in particular
accretion-powered X-ray sources. ChaMPlane
\citep{grindlay2003,grindlay05} comprises two phases of study of the
Galactic plane (Galactic latitudes $|$b$|$ $<$ 12$\degr$): firstly an
X-ray survey of serendipitous sources from archival deep
$\it{Chandra}$ X-ray Observatory pointings (with exposure times
greater than $\sim$20ks), and secondly an optical survey in $H\alpha$
(narrow band) and Johnson $V$, $R$, and $I$ filters, using the Mosaic
camera on the CTIO and KPNO 4m telescopes to image 36$'$ $\times$
36$'$ fields centered on the $\it{Chandra}$ pointings. Optical spectra
are then obtained for classification of candidate optical counterparts
to X-ray sources. Infrared (IR) observations are used to identify
candidate counterparts in heavily obscured fields.

In this paper we analyze a sample of candidate optical counterparts
from five ChaMPlane fields towards the Galactic bulge using
low-resolution optical spectra. These fields are within 20$\degr$ of
the Galactic Center (GC), and within 3$\degr$ of the Galactic
plane. Using the optical band necessarily constrains the scope of this
project: stars at the distance of the GC ($\sim$8 kpc) are hidden
behind a hydrogen column density $N_H \sim$ 0.5--2.0 $\times$
10$^{23}$ cm$^{-2}$, and are thus absorbed by $A_V$ $\gtrsim$ 25
($A_R$ $\gtrsim$ 19). Given our optical survey limit of $R$ = 24 and
this level of extinction, optical counterparts at the GC are
unobservable, therefore our work is restricted to foreground (d
$\lesssim$ 3 kpc) sources. As a consequence we focus our efforts on
two main goals: 1) to identify candidate cataclysmic variables (CVs)
through their broad $H\alpha$ line emission and 2) to study the
properties of the sample of stellar coronal emission sources that we
detect. In doing so we highlight anomalous cases as potential active
binary or quiescent low-mass X-ray binary (qLMXB) systems.

The X-ray and optical datasets used in this study, and their
reduction, are discussed in $\S$2--3. In $\S$3 we highlight new
techniques developed for our analysis---a simple spectral fitting
process to obtain the extinction $E(B-V)$ and thus $N_H$ from the
optical spectra, a three-dimensional (3D) dust model of the Galaxy
\citep{drimmel01} to obtain a distance in any direction given
$E(B-V)$, and the X-ray Quantile Color-Color Diagram (QCCD) technique
for spectral analysis of low-count X-ray point sources. In $\S$4 we
give results for X-ray and optical source properties and present our
detected candidate CVs and observed stellar X-ray luminosity
function. In $\S$5 we discuss the likely composition of our stellar
coronal source population and compare the CV sample to predictions of
a simple Galactic distribution model. Finally in $\S$6 we present our
conclusions.

\section{Observations and Datasets}

\subsection{X-ray Dataset}
\citet{hong05} and Hong et al. (2008, in preparation) describe in
detail our pipeline processing of archival {\it Chandra} observations
for use in the ChaMPlane survey and subsequent, more detailed
analysis. In summary, source lists from detections in a broad (B$_X$,
0.3--8.0 keV), soft (S$_X$, 0.3--2.5 keV) and hard (H$_X$, 2.5--8.0
keV) energy band, are cross-correlated to form a master source
list. Source properties such as flux and energy quantiles (see $\S
3.1$), are derived in energy bands more appropriate to analysis of low
count sources. These conventional bands are defined: S$_C$ (0.5--2.0
keV), H$_C$ (2.0--8.0 keV) and B$_C$ (0.5--8.0 keV).  In the following
analysis, we consider only X-ray sources of level 1 and above---these
being sources unaffected by hot pixels, bad columns or bad bias values
on the ACIS detector, or readout streaks from bright sources as well
as sources too close to the chip boundary; see \citet{hong05} for a
complete description of the different levels assigned to sources in
our X-ray catalog.

\begin{deluxetable}{ccrrcccr}
\tablenum{1} \tablewidth{0pt} \tabletypesize{\scriptsize}
\tablecaption{X-ray Observations used in this Paper}
\tablehead{\colhead{Obs. ID} & \colhead{Field Name} &
\multicolumn{2}{c}{Aimpoint}&\colhead{No. of
Sources\tablenotemark{a}}&\colhead{Exposure}&\colhead{Effective exp.}&\colhead{$N_{22}$}\\
\colhead{} & \colhead{} & \colhead{$\it{l}$(\degr)} &
\colhead{$\it{b}$(\degr)} &\colhead{B$_{X}$} & \colhead{(ks)} & \colhead{(ks)} &
\colhead{(cm$^{-2}$)\tablenotemark{b}}} \startdata 
99 & GRO J1655$-$40 (J1655) & 344.98178 & 2.45612 & 137 & 43.0 & 42.3 & 1.0\\ 
737 & G347.5$-$0.5b (G347b) & 347.36606 & $-$0.85734 & 108 & 40.0 & 38.2 & 3.3\\ 
944 & SgrB2 & 0.58834 & $-$0.02491 & 369 & 100.0 & 96.8 & 10.8\\ 
945 & Gal. Center Arc (GalCA) & 0.14055 & $-$0.09707 & 222 & 50.0 & 48.5 & 10.8\\ 
53392\tablenotemark{c} & SgrA$\star$ & 359.94415 & $-$0.04594 & 2982 & 748.0 & 698.0 & 4.3 
\enddata
\tablenotetext{a}{\scriptsize{Number of valid (level 1) sources found
in the B$_X$ band.}}  \tablenotetext{b}{\scriptsize{Integrated column
density $N_H$ in units of 10$^{22}$ up to a distance of 8 kpc, from
\citet{drimmel01}.}}  \tablenotetext{c}{\scriptsize{ObsID 53392 is a
number we assign internally to refer to our stacked SgrA$\star$
observation.}}
\end{deluxetable}

X-ray data for this paper come from three ACIS-I and one ACIS-S
$\it{Chandra}$ pointings. In addition, we have stacked 14 ACIS-I
observations centered on SgrA* to create a deep image of the Galactic
Center region. This stack includes the observations analyzed by
\citet{muno} (with the exception of ObsID 1561), that amount to a
total of 590 ksec exposure time, as well as ObsIDs 3549, 4683, 4684
and 5360. The resultant total exposure time is 748ks or 698ks of good
time after processing. The process of dealing with duplicate sources
between individual pointings in the stack is detailed in
\citet{hong05}.  The final stacked SgrA* observation we label ObsID
53392. It overlaps partially with ObsID 945---the two share 46 X-ray
sources in common. All fields listed in Table 1 were observed with
$\it{Chandra}$ ACIS-I except J1655, for which ACIS-S was
used. Exposure times are given before and after correcting for good
time intervals (GTI).

\subsection{Optical Dataset}
Targets for optical spectral follow-up were selected following
observations made in March 2000 with the Mosaic camera on the CTIO 4m
telescope in $V$, $R$, $I$ and $H\alpha$ filters to identify candidate
counterparts as follows \citep[see][for details]{zhao05}. We determine
the systematic offsets between the Chandra and Mosaic astrometries,
i.e. the boresight correction, for each pair of X-ray and optical
images using the iterative procedure described in
\citet{zhao05}. After applying the boresight correction, we search for
candidate optical counterparts within some confidence radius of each
X-ray source, taking into account optical and X-ray astrometric
uncertainties and the boresight error. We elected to search within a 2
$\sigma$ radius of each source (thus losing on average $\sim$5\% of
the real counterparts). Table 2 summarizes these results. The
resultant catalog of candidate counterparts was used to make a target
list for the observing runs in 2001--2002 with the Low Dispersion
Survey Spectrograph 2 (LDSS2) on the 6.5m Baade Magellan telescope.

\begin{deluxetable}{lccccc}
\tabletypesize{\scriptsize} \tablenum{3} \tablewidth{0pt}
\tablecaption{X-ray to Optical Matching Summary}
\tablehead{\colhead{Field Name} &
\colhead{N(X)\tablenotemark{a}} & \colhead{N(Xmatch)\tablenotemark{b}}
& \colhead{N(Opt)\tablenotemark{c}} &
\colhead{N(Spectra)\tablenotemark{d}} &
\colhead{N(Id. Spec.)\tablenotemark{e}}} \startdata 
J1655 & 137 & 58 & 227 & 32 & 28 \\ 
G347b & 108 & 53 & 147 & 26 & 25 \\ 
SgrB2 & 369 & 112 & 165 & 52 & 45 \\ 
GalCA & 222 & 48 & 67 & 9 & 9 \\ 
SgrA$\star$ & 2982 & 327 & 340 & 17 & 15 \\
\hline Total & 3818 & 598 & 946 & 136 & 122
\enddata
\tablecomments{$^{a}$The number of unique (level 1) X-ray sources in
all bands. $^{b}$The number of X-ray sources that have any
optical counterpart(s). $^{c}$The number of optical sources falling
inside 2$\sigma$ error circles. Some X-ray sources match multiple
optical sources. $^{d}$The number of optical spectra of $\it{Chandra}$
matches obtained with LDSS2. $^{e}$The number of $\emph{identifiable}$
spectra from this sample.}
\end{deluxetable}

The LDSS2
instrument\footnote{http://www.ociw.edu/lco/magellan/instruments/LDSS2/ldss2\_usersguide.html}
uses a multi-aperture mask with a $\sim$5$'$ diameter field of
view. With a dispersion of 5.3$\textrm{\AA}$ per pixel, we obtained
spectra centered on 6500$\textrm{\AA}$ covering $\sim$3500 to
9500$\textrm{\AA}$ at a resolution of 13.3$\textrm{\AA}$.  Masks for
LDSS2 were generated with the $\emph{ldss2mask.f}$ FORTRAN code. Using
slit lengths between $\sim$5 and 10$''$, and 3 or 4 alignment stars,
between 6 and $\sim$20 targets were assigned to each mask. Given the
ChaMPlane goal of finding X-ray binaries largely powered by accretion
processes, highest priority for inclusion of targets on the masks went
to objects showing both X-ray and $H\alpha$ emission (i.e. $H\alpha -
R < -$0.3), followed by X-ray source candidate optical counterparts
(regardless of X-ray or optical colors), then $H\alpha$ bright objects
($H\alpha - R < -$0.3), and then `marginal' $H\alpha$ objects (with
$-$0.3 $< H\alpha - R < -$0.2). Each of these groups was sorted in
order of R-magnitude brightness, with brightest objects having highest
priority. Details of the observations are given in Table 3. 

Data from LDSS2 were reduced using the standard IRAF procedure
$\emph{ccdproc}$. Spectra were extracted one-by-one using the IRAF
$\emph{doslit}$ package. When crowding in dense fields resulted in
multiple stars falling on a slit, the correct target star was
identified for extraction by comparing the dispersed CCD image of the
spectra with the Mosaic image of the field and a reference image of
the sky taken without the LDSS2 slit mask and grism in place. Stars
too poorly exposed to find any trace on the CCD, too badly saturated,
or on incorrectly cut slits were not extracted (these three cases
account for $\sim$20\% of the original targets). In total we extracted
136 usable targets within the five $\it{Chandra}$ fields-of-view. Flux
calibration was performed on each extracted spectrum using IRAF
routines with flux standard spectra taken each night. Where possible,
all extracted spectra were then assigned a spectral classification by
visual inspection and comparison with published atlases of optical and
near-IR spectra \citep{dodgen93,jacoby84,carquill97,andrill95}.

\begin{deluxetable}{lcccc}
\tablenum{2} 
\tablewidth{0pt} 
\tabletypesize{\scriptsize}
\tablecaption{Spectral Observations with LDSS2/Magellan 6.5m}
\tablehead{ \colhead{Date} & \colhead{Fields Observed} & \colhead{Flux
Standards Used} & \colhead{Grism/Filter\tablenotemark{a}}}
\startdata 
May 18--20 2001 & SgrA & LTT3864, LTT7987 & MedRed/S2\\ 
July 25--27 2001 & SgrB2 & LTT9239, Feige110, LTT7379 & MedBlue/None\\ 
June 16--19 2002 & J1655, GalCA, SgrA, SgrB2, G347b & LTT9239, EG274 & MedBlue/None
\enddata
\tablenotetext{a}{\scriptsize{2$^{nd}$ order blocking filter}}
\end{deluxetable}

\section{Analysis}
\subsection{Extinction and Hydrogen Column Density}
To estimate the hydrogen column density $N_H$ for all X-ray-optical
matches observed with LDSS2, we use the flux-calibrated {\it optical}
spectrum of each candidate optical counterpart. We fit the spectrum
with a blackbody attenuated by interstellar reddening, using the
analytical parameterization of the average Galactic extinction law
$E(\lambda -V)/E(B-V)$ given by \citet{howarth83}. We limit the region
to be fitted to the central part of the spectra:
5500--6700$\textrm{\AA}$; in part to simplify the fitting procedure,
and also because the flux calibrated spectra anomalously fall off in
flux redward of $\sim$7000$\textrm{\AA}$. We also interpolate over
major spectral lines at 5575, 6300 and 6562$\textrm{\AA}$ and in the
case of the molecular bands present in M type stars we fit a smooth
continuum shape to the star at 3--4 points between bands (we use a
Legendre polynomial of order 10) and fit to this instead. Using a
fixed effective temperature (T$_{eff}$), estimated from our spectral
type determination from spectral lines, we fit for $E(B-V)$ and a
normalization factor: $R^2$/$d^2$ (radius of object $\it{R}$, distance
$\it{d}$). We obtain $A_V$ from the standard relation $A_V$ = 3.1
$\times E(B-V)$ and the hydrogen column density from $N_H$ = 1.79
$\times$ 10$^{21}$ $\times A_V$ cm$^{-2}$
\citep[from][]{predehl95}. The T$_{eff}$ in each case is
representative of the spectral type range we were able to estimate
given only visual classification of each spectrum and no constraint on
luminosity class (i.e. A/F means A8 to F2, luminosity class
undetermined). The fit is independent of any X-ray emission
properties. Effective temperatures and adopted photometry for each
spectral type `bin' are given in Table 4 below, (see references listed
in the table). Example fits for a K and an M (polynomial fit) star are
shown in Figure 1.

To test the above, we derive comparison $N_H$ values using our $V$,
$R$ and $I$ photometry (where available). We calculate $E(V-R)$
$\equiv$ $V-R - (V-R)_0$ and $E(R-I)$ $\equiv$ $R-I - (R-I)_0$,
selecting appropriate values for the intrinsic ($V-R)_0$ and ($R-I)_0$
colors \citep[from][]{Allen2000} for a given spectral type (see Table
4). From \citet{dopita03} we find: $A_V$ = 3.97 $\times E(V-R)$ and
$A_V$ = 3.76 $\times E(R-I)$ and we convert to $N_H$ as above. Figure
2 below shows spectral fit-determined values of $N_H$ against the
color determined results. The values are correlated somewhat, but the
spectral fit $N_H$ is systematically higher than that from photometry:
from a simple linear regression test we find $N_H(fit)$ =
0.96$\pm$0.09($N_H(R-I)$) + 0.2 and $N_H(fit)$ =
0.81$\pm$0.07($N_H(V-R)$) + 0.25. Dropping the five highest points in
either plot however, worsens the correlation: $N_H(fit)$ =
0.81$\pm$0.13($N_H(R-I)$) + 0.27 and $N_H(fit)$ =
0.65$\pm$0.1($N_H(V-R)$) + 0.3. This overestimate in $N_H$ is also
seen in a color-magnitude diagram for our stellar sources (Figure 3)
of absolute visual magnitude $M_V$ against dereddened color $V-R$. A
comparison main sequence is plotted using the data from Table 4. We
estimate the systematic excess in $A_V$ from our fitting technique to
be 1.1 (equivalent to $N_H$ of $2\times 10^{21}$cm $^{-2}$). This
overestimate is perhaps due to the inadequacy of the black-body
spectrum as a fit to late-type stellar spectra over this wavelength
range where many absorption lines significantly modify the continuum
shape.

\begin{deluxetable}{lccccc}
\tabletypesize{\scriptsize} \tablenum{4} \tablewidth{0pt}
\tablecaption{Effective Temperature Assignment and Photometric Properties
Used} \tablehead{ \colhead{Classification} & \colhead{Sp. Type Range}
& \colhead{T$_{eff}$ (K)} & \colhead{V$-$R} & \colhead{R$-$I} &
\colhead{M$_V$}} \startdata 
Early A & A2--A4 & 9100 & 0.08 & 0.03 & 1.8$\pm$0.4\\ 
Mid A & A4--A6 & 8500 & 0.16 & 0.06 & 2.0$\pm$0.5\\ 
Late A & A6--A8 & 7700 & 0.19 & 0.09 & 2.4$\pm$0.6\\ 
A/F & A8--F2 & 7200 & 0.30 & 0.17 & 2.8$\pm$0.6\\ 
Early F & F2--F4 & 6750 & 0.35 & 0.20 & 3.2$\pm$0.4\\ 
Mid F & F4--F6 & 6500 & 0.40 & 0.24 & 3.5$\pm$0.4\\ 
Late F & F6--F8 & 6200 & 0.47 & 0.29 & 3.8$\pm$0.4\\ 
F/G & F8--G2 & 5950 & 0.50 & 0.31 & 4.3$\pm$0.7\\ 
Early G & G2--G4 & 5600 & 0.53 & 0.33 & 4.6$\pm$0.4\\ 
Mid G & G4--G6 & 5400 & 0.60 & 0.42 & 4.9$\pm$0.5\\ 
Late G & G6--G8 & 5200 & 0.64 & 0.43 & 5.3$\pm$0.5\\ 
G/K & G8--K2 & 5050 & 0.70 & 0.48 & 5.8$\pm$1.0\\ 
Early K & K2--K4 & 4700 & 0.80 & 0.53 & 6.7$\pm$0.6\\ 
Mid K & K4--K6 & 4300 & 1.10 & 0.75 & 7.4$\pm$0.6\\ 
Late K & K6--K8 & 4000 & 1.15 & 0.78 & 8.4$\pm$0.6\\ 
K/M & K8--M2 & 3700 & 1.25 & 0.93 & 9.3$\pm$1.0\\ 
Early M & M2--M4 & 3400 & 1.42 & 1.15 & 11.1$\pm$1.6\\ 
Mid M & M4--M6 & 3000 & 1.8 & 1.67 & 13.2$\pm$2.0
\enddata \tablecomments{References for colors, temperatures and
absolute magnitudes: \citet{johnson66}, \citet{bessell91},
\citet{mikami82}, \citet{houk97}, \citet{Gray1992},
\citet{Allen2000}. We estimate an uncertainty in the assigned
T$_{eff}$ values of $\pm$500K, based on the spread of values found in
the various references.}
\end{deluxetable}

\begin{deluxetable}{lllrcccccccccc}
\tablecolumns{13}
\tabletypesize{\tiny}
\rotate
\tablenum{5}
\tablewidth{0pt}
\tablecaption{Combined Results for 2$\sigma$ Optical Matches}
\tablehead{\colhead{ChOPS ID\tablenotemark{a}} & \colhead{Type\tablenotemark{b}} & \colhead{Special\tablenotemark{c}} & \colhead{Counts\tablenotemark{d}} & \colhead{$N_{22}$\tablenotemark{e}} & \colhead{$\log(F_x/F_R)$\tablenotemark{f}} & \colhead{Dist.\tablenotemark{g}} & \colhead{M$_V$} & \colhead{$\log(L_x)$\tablenotemark{h}} & \colhead{R} & \colhead{N$_{M}$\tablenotemark{i}} & \colhead{Srch\tablenotemark{j}} & \colhead{Offset\tablenotemark{j}} & \colhead{P$_{Rn}$\tablenotemark{j}}\\
\colhead{(1)} & \colhead{(2)} &\colhead{(3)} &\colhead{(4)} &  \colhead{(5)} & \colhead{(6)} & \colhead{(7)} & \colhead{(8)} & \colhead{(9)} &\colhead{(10)} & \colhead{(11)} & \colhead{(12)} & \colhead{(13)}  & \colhead{(14)}} 
\startdata 
\sidehead{J1655}
J165343.19$-$394826.1 & G/K & \nodata & 11$\pm$5 & 0.3(1) & $-$2.5$\pm$0.4 & 1.4 & 5.7(9) & 29.54$\pm$0.7 & 17.48(1) & 1 & 1.199 & 0.420 & 0.207\\
J165350.37$-$394621.9 & midM dMe? & \nodata & 12$\pm$5 & 0.2(2)  & $-$1.8$\pm$0.5 & 1.4 & 8(2) & 29.43$\pm$0.8 & 19.32(1) & 2 & 1.886 & 1.271 & 0.712\\
J165350.70$-$395339.6 & earlyK & \nodata & 8$\pm$5 & 0.3(1) & $-$2.2$\pm$0.4 & 1.5 & 6.8(9) & 29.46$\pm$0.7 & 18.44(1) & 1 & 0.858 & 0.788 & 0.082\\
J165351.90$-$394812.7 & F/G? & $by$ & 18$\pm$6 & 0.7(2) & $-$1.8$\pm$0.6 & 4.0 & 5(2) & 30.59$\pm$0.8 & 20.63(1) & 1 & 0.715 & 0.692 & 0.064\\
J165354.54$-$394741.0 & G? & $rs$ & 11$\pm$5 & 0.9(2) & $-$3.5$\pm$0.7 & 4.5 & 1(1) & 30.38$\pm$5.9 & 17.96(1) & 1 & 0.874 & 0.322 & 0.092\\
J165356.10$-$394732.7 & lateF & \nodata & 9$^{\star}\pm$5 & 0.4(2) & $-$3.1$\pm$0.5 & 1.9 & \nodata & 29.87$\pm$0.7 & 16.30(1) & 1 & 0.933 & 0.391 & 0.113\\
J165356.98$-$394636.7 & earlyG & \nodata & 8$^{\star}\pm$5 & 0.9(2) & $-$2.8$\pm$0.6 & 4.8 & 2(2) & 30.81$\pm$5.4 & 19.10(3) & 1 & 1.322 & 0.358 & 0.271\\
J165357.34$-$394847.9 & midM dMe? & \nodata & 18$\pm$5 & 0.2(1) & $-$1.3$\pm$0.4 & 1.1 & 9.9(9) & 29.43$\pm$0.7 & 19.93(2) & 1 & 0.677 & 0.380 & 0.050\\
J165357.85$-$395032.1 & F/G? & \nodata & 6$^{\star}\pm$4 & 0.6(1) & $-$2.5$\pm$0.5 & 3.1 & 4.1(9) & 30.12$\pm$0.8 & 19.20(1) & 1 & 0.584 & 0.230 & 0.024\\
J165358.28$-$394812.6 & midM & \nodata & 6$^{\star}\pm$4 & 0.1(1) & $-$2.1$\pm$0.5 & 0.8 & 10.0(8) & 28.50$\pm$0.8 & 19.25(1) & 1 & 1.045 & 0.510 & 0.163\\
J165358.44$-$394745.6 & early/midM & \nodata & 15$\pm$5 & 0.2(2) & $-$2.4$\pm$0.6 & 1.2 & 7(2) & 29.46$\pm$0.7 & 17.45(1) & 1 & 0.842 & 0.211 & 0.081\\
J165359.83$-$394856.1 & early/midM dMe & \nodata & 9$\pm$4 & 0.3(3) & $-$1.8$\pm$0.6 & 1.4 & \nodata & 29.40$\pm$0.8 & 19.58(2) & 1 & 0.698 & 0.399 & 0.052\\
J165400.14$-$395045.1 & F3--6IV & $q$ & 64$\pm$9 & 0.6(1) & $-$2.9$\pm$0.4 & 2.8 & 1.2(9) & 30.86$\pm$0.7 & 16.14(1) & 1 & 0.412 & 0.218 & 0.015\\
J165400.81$-$395145.5 & earlyM dMe?  \nodata && 24$\pm$6 & 0.4(1) & $-$0.9$\pm$0.4 & 2.0 & \nodata & 30.04$\pm$0.7 & 21.50(5) & 1 & 0.444 & 0.237 & 0.013\\
J165402.81$-$395104.8 & earlyG & \nodata & 8$^{\star}\pm$4 & 0.4(2) & $-$2.7$\pm$0.5 & 2.2 & 4(2) & 29.88$\pm$0.8 & 17.77(1) & 1 & 0.511 & 0.268 & 0.017\\
J165403.00$-$394816.3 & midM dMe & \nodata & 17$\pm$5 & 0.2(1) & $-$1.2$\pm$0.4 & 1.1 & 10.6(9) & 29.26$\pm$0.7 & 20.56(1) & 1 & 0.711 & 0.360 & 0.055\\
J165403.27$-$395217.2 & midlateG & \nodata & 14$\pm$5 & 0.4(4) & $-$2.5$\pm$0.8 & 2.1 & 4(3) & 30.08$\pm$0.8 & 17.58(1) & 1 & 0.484 & 0.275 & 0.017\\
J165403.64$-$395054.7 & lateG & \nodata & 23$\pm$6 & 0.3(1) & $-$2.2$\pm$0.4 & 1.6 & 5.6(9) & 29.90$\pm$0.7 & 17.77(1) & 1 & 0.456 & 0.197 & 0.015\\
J165404.79$-$394817.4 & G/K( & $rs$ & 21$\pm$6 & 0.8(2) & $-$2.7$\pm$0.5 & 4.0 & 1(1) & 31.05$\pm$1.2 & 17.64(1) & 1 & 0.636 & 0.451 & 0.046\\
J165404.89$-$394934.5 & K/M & \nodata & 9$\pm$4 & 0.2(1) & $-$2.2$\pm$0.4 & 0.9 & 8.1(9) & 28.99$\pm$0.7 & 18.19(2) & 1 & 0.625 & 0.474 & 0.036\\
J165407.44$-$394542.7 & ? & \nodata & 13$\pm$6 & $<$3.5 & $> -$1.3 & $<$2.4 & $>$7.43 & $<$30.2 & 20.87(1) & 3 & 2.172 & 1.383 & 0.950\\
J165408.14$-$395636.1 & midG & $q$ & 86$\pm$10 & 0.4(1) & $-$2.7$\pm$0.5 & 2.2 & 3(1) & 31.00$\pm$0.7 & 16.19(2) & 1 & 0.724 & 0.170 & 0.056\\
J165411.04$-$395200.6 & midM? dMe & \nodata & 9$\pm$4 & 0.5(1) & $-$1.3$\pm$0.4 & 2.3 & 8.0(9) & 30.03$\pm$0.7 & 21.14(2) & 1 & 0.635 & 0.105 & 0.029\\
J165411.38$-$395236.8 & early/midG & $rs$ & 13$\pm$5 & 0.8(2) & $-$2.8$\pm$0.5 & 3.9 & \nodata & 31.00$\pm$1.1 & 17.51(1) & 1 & 0.601 & 0.380 & 0.032\\
J165412.28$-$395434.2 & mid/lateG & \nodata & 41$\pm$8 & 0.3(2) & $-$2.6$\pm$0.5 & 1.7 & 3(2) & 30.43$\pm$0.7 & 15.75(1) & 1 & 0.668 & 0.232 & 0.042\\
J165413.21$-$395005.1 & early/midK & \nodata & 6$^{\star}\pm$4 & 0.2(1) & $-$2.8$\pm$0.4 & 1.3 & 6.1(9) & 29.18$\pm$0.7 & 17.25(1) & 1 & 0.814 & 0.169 & 0.072\\
J165413.51$-$394757.0 & mid/lateF & \nodata & 7$^{\star}\pm$4 & 0.2(1) & $-$3.4$\pm$0.4 & 1.3 & \nodata & 29.15$\pm$0.7 & 15.92(1) & 1 & 1.152 & 0.210 & 0.183\\
J165413.47$-$395757.6 & ? & \nodata & 22$\pm$6 & $<$0.5$\dagger$ & $> -$1.0 & $<$2.5 & $>$7.86 & $<$30.5 & 21.33(3) & 1 & 1.647 & 1.257 & 0.406\\
J165419.80$-$395445.8 & ? & \nodata & 11$\pm$5 & $<$0.01$\dagger$ & $> -$1.7 & $<$0.1 & $>$15.7 & $<$27.0 & 19.62(1) & 1 & 1.673 & 0.650 & 0.511\\
J165422.01$-$395205.0 & midM dMe & \nodata & 12$\pm$5 & 0.2(2) & $-$1.5$\pm$0.5 & 1.3 & 9(2) & 29.58$\pm$0.7 & 19.57(1) & 1 & 0.972 & 0.920 & 0.087\\
J165424.36$-$395141.0 & early/midK & \nodata & 10$\pm$5 & 0.04(14) & $-$3.2$\pm$0.4 & 0.2 & \nodata & 27.85$\pm$0.8 & 15.32(1) & 1 & 1.519 & 0.217 & 0.317\\
J165427.11$-$395231.4 & ? & \nodata & 21$\pm$6 & $<$1.0 & $> -$1.9 & $<$6.3 & $>$1.71 & $<$31.7 & 20.01(1) & 1 & 1.500 & 0.208 & 0.324\\
\sidehead{G347b}
J171447.10$-$395247.1 & lateG & \nodata & 25$\pm$8 & 0.6(1)  & $-$2.6$\pm$0.5 & 2.4 & 4.0(9) & 30.32$\pm$0.7 & 18.22(2) & 1 & 2.190 & 0.790 & 0.514\\
J171448.72$-$395617.3 & G? & $rs$ & 9$^{\star}\pm$5 & 1.3(2) &  $-$2.7$\pm$0.6 & 4.2 & 2(1) & 30.98$\pm$0.7 & 20.20(1) & 1 & 2.043 & 1.002 & 0.456\\
J171452.55$-$395640.0 & lateG & \nodata & 22$\pm$6 & 0.6(1) &  $-$2.4$\pm$0.4 & 2.3 & 3.8(9) & 30.53$\pm$0.6 & 17.95(1) & 1 & 0.961 & 0.173 & 0.104\\
J171503.40$-$395416.2 & lateF & \nodata & 11$\pm$5 & 0.6(1) & $-$2.2$\pm$0.5 & 2.0 & 5.7(9) & 29.96$\pm$0.7 & 19.25(2) & 1 & 1.011 & 0.475 & 0.116\\
J171507.76$-$395407.5 & ? & \nodata & 8$^{\star}\pm$4 & $>$10.0 & \nodata & \nodata & \nodata & \nodata & 21.85(3) & 1 & 1.170 & 0.650 & 0.168\\
J171508.43$-$395530.9 & earlyK & $rs$ & 10$\pm$5 & 1.1(1) & \nodata & 3.7 & 4(1) & 30.94$\pm$0.6 & 21.03(5) & 1 & 0.756 & 0.256 & 0.069\\
J171509.91$-$395429.6 & lateF? & \nodata & 8$^{\star}\pm$4 & 0.9(1) & $-$2.2$\pm$0.5 & 3.4 & 4(1) & 30.63$\pm$0.6 & 19.82(1) & 1 & 1.128 & 0.555 & 0.158\\
J171510.54$-$395603.5 & earlyA & \nodata & 13$\pm$5 & 0.5(2) & $-$3.4$\pm$0.6 & 2.1 & \nodata & 30.34$\pm$0.6 & 14.96(1) & 1 & 0.655 & 0.639 & 0.049\\
J171513.56$-$395456.6 & G/K? & $tt$ &12$\pm$5 & 0.6(1) & $-$2.5$\pm$0.4 & 2.7 & 3.4(9) & 30.61$\pm$0.6 & 18.00(1) & 1 & 0.760 & 0.425 & 0.068\\
J171513.65$-$400112.3 & mid/lateG & \nodata & 20$\pm$6 & 0.5(2) & $-$2.1$\pm$0.5 & 2.1 & \nodata & 30.55$\pm$0.6 & 17.81(1) & 1 & 0.653 & 0.371 & 0.048\\
J171513.90$-$395743.4 & earlyG & \nodata & 10$\pm$5 & 0.6(1) & $-$2.6$\pm$0.5 & 2.7 & 4.0(9) & 30.22$\pm$0.6 & 18.32(1) & 1 & 0.517 & 0.299 & 0.027\\
J171516.16$-$395844.1 & midG & \nodata & 8$^{\star}\pm$4 & 0.7(2) & $-$3.1$\pm$0.6 & 3.0 & 2(2) & 30.58$\pm$0.6 & 16.952(4) & 1 & 0.571 & 0.474 & 0.026\\
J171517.18$-$395535.7 & F/G? & \nodata & 9$\pm$4 & 0.8(1) & $-$2.3$\pm$0.6 & 3.0 & 4(1) & 30.36$\pm$0.7 & 19.73(1) & 1 & 0.692 & 0.337 & 0.059\\
J171518.23$-$400147.9 & K/M dMe& \nodata & 12$\pm$5 & 0.5(1) & $-$1.4$\pm$0.4 & 2.2 & 8(1) & 30.19$\pm$0.6 & 20.46(1) & 1 & 1.386 & 1.020 & 0.224\\
J171519.78$-$400411.6 & G/K? & \nodata & 12$^{\star}\pm$6 & 0.8(2) & $-$2.5$\pm$0.5 & 3.0 & 3(1) & 30.81$\pm$0.6 & 17.995(4) & 1 & 1.630 & 1.219 & 0.271\\
J171520.17$-$400248.2 & G? & $by$ & 56$\pm$9 & 0.6(2) & \nodata & 2.5 & 6(1) & 31.04$\pm$0.5 & 20.05(3) & 1 & 0.691 & 0.343 & 0.054\\
J171520.66$-$395934.9 & F/G & \nodata & 5$^{\star}\pm$4 & 0.5(1) & $-$3.1$\pm$0.5 & 2.2 & 3(1) & 30.05$\pm$0.6 & 16.86(1) & 2 & 0.793 & 0.354 & 0.077\\
J171523.89$-$400026.8 & lateK & $rs$ & 13$\pm$5 & 0.7(2) & $-$2.6$\pm$0.5 & 2.9 & 3(2) & 30.82$\pm$0.6 & 16.80(1) & 1 & 0.577 & 0.556 & 0.040\\
J171524.21$-$395950.9 & lateG & $rs$ & 17$\pm$5 & 0.7(2) & $-$2.8$\pm$0.5 & 2.8 & 1(1) & 30.98$\pm$0.6 & 16.83(1) & 1 & 0.644 & 0.364 & 0.048\\
J171525.99$-$395603.9 & lateG? & \nodata & 18$\pm$6 & 1.0(6) & $-$3.0$\pm$1.2 & 3.6 & 1(5) & 31.03$\pm$0.7 & 17.93(1) & 1 & 0.609 & 0.535 & 0.042\\
J171528.22$-$395929.4 & midG & \nodata & 9$\pm$5 & 0.6(1)  & $-$2.2$\pm$0.5 & 2.7 & 4.7(9) & 30.33$\pm$0.6 & 19.23(1) & 1 & 0.768 & 0.437 & 0.063\\
J171529.77$-$400309.2 & lateA & \nodata & 14$\pm$6 & 0.4(2) & $-$3.1$\pm$0.5 & 1.8 & 3(2) & 30.00$\pm$0.6 & 16.06(2) & 1 & 2.004 & 1.086 & 0.404\\
J171531.96$-$395259.3 & midG & $by$ & 18$\pm$6 & 0.7(1) & $-$1.1$\pm$0.4 & 3.0 & 7(1) & 30.94$\pm$0.6 & 21.18(3) & 1 & 1.343 & 0.678 & 0.189\\
J171532.99$-$400220.4 & midM & \nodata & 19$\pm$7 & 0.2(2) & $-$2.1$\pm$0.5 & 1.1 & 7(2) & 29.71$\pm$0.6 & 17.317(4) & 1 & 1.100 & 0.770 & 0.114\\
J171536.42$-$395532.7 & earlyG & \nodata & 11$\pm$5 & 0.5(1) & $-$2.7$\pm$0.4 & 2.1 & 3.6(9) & 30.25$\pm$0.6 & 17.08(1) & 1 & 1.079 & 0.288 & 0.099\\
J171537.99$-$395739.2 & early/midG & \nodata & 14$\pm$5 & 0.6(1) & $-$2.1$\pm$0.4 & 2.5 & 4.9(9) & 30.44$\pm$0.6 & 18.74(1) & 1 & 0.934 & 0.418 & 0.063\\
\sidehead{SgrB2}
J174633.29$-$282512.7 & mid/lateF & \nodata & 25$\pm$9 & 0.8(2) & $-$3.3$\pm$0.5 & 1.3 & 3(1) & 30.07$\pm$1.3 & 16.52(1) & 2 & 2.017 & 1.464 & 0.282\\
J174634.65$-$282721.1 & ? & \nodata & 33$\pm$8 & $>$2.0 & $< -$3.2 & $>$2.3 & $< -$0.09 & $>$30.7 & 21.18(2) & 3 & 3.641 & 3.440 & 1.749\\
J174639.48$-$282804.3 & G/K & \nodata & 24$\pm$7 & 0.6(2) & $-$2.7$\pm$0.5 & 1.0 & 5(2) & 29.64$\pm$1.5 & 17.667(4) & 1 & 1.146 & 0.097 & 0.086\\
J174640.67$-$282417.1 & F/G & \nodata & 25$\pm$7 & 0.4(2) & $-$3.7$\pm$0.5 & 0.7 & 4(2) & 29.24$\pm$1.6 & 14.53(1) & 1 & 1.213 & 1.153 & 0.100\\
J174642.22$-$282907.4 & K & \nodata & 21$\pm$7 & 2.5(4) & $-$5.2$\pm$0.9 & 2.6 & $-$4(3) & 30.88$\pm$1.2 & 18.08(1) & 1 & 1.171 & 0.885 & 0.072\\
J174645.48$-$282926.2 & F/G & \nodata & 225$\pm$16 & 0.5(2) & $-$3.3$\pm$0.5 & 0.9 & 1(1) & 30.47$\pm$1.5 & 13.54(2) & 1 & 0.561 & 0.419 & 0.018\\
J174647.09$-$282426.1 & mid/lateG & \nodata & 12$^{\star}\pm$6 & 0.6(2) & $-$4.1$\pm$1.9 & 0.9 & 6(1) & 28.10$\pm$2.4 & 17.506(2) & 1 & 1.108 & 0.696 & 0.087\\
J174649.08$-$282645.6 & G/K & \nodata & 11$\pm$5 & 0.7(3) & $-$2.7$\pm$0.6 & 1.0 & 6(2) & 29.37$\pm$1.5 & 18.52(1) & 1 & 1.015 & 0.749 & 0.063\\
J174649.59$-$282338.1 & midM dMe& \nodata & 13$\pm$6 & 0.3(3) & $-$2.1$\pm$0.6 & 0.5 & 10(2) & 28.53$\pm$1.7 & 18.82(1) & 1 & 2.399 & 0.899 & 0.917\\
J174650.65$-$282839.8 & lateF & \nodata & 32$\pm$7 & 0.6(2) & $-$3.3$\pm$0.5 & 1.0 & 3(1) & 29.74$\pm$1.5 & 15.75(9) & 1 & 0.687 & 0.110 & 0.033\\
J174651.71$-$283030.2 & midM dMe& \nodata & 40$\pm$8 & 0.1(2) & $-$1.9$\pm$0.5 & 0.2 & 12(2) & 28.17$\pm$2.2 & 18.04(2) & 1 & 0.787 & 0.410 & 0.051\\
J174652.63$-$282503.1 & F/G? & \nodata & 33$\pm$7 & 0.6(2) & $-$2.8$\pm$0.5 & 0.9 & 5(1) & 29.69$\pm$1.5 & 16.709(4) & 1 & 0.647 & 0.550 & 0.031\\
J174653.22$-$282830.9 & lateG & $rs$ & 209$\pm$16 & 0.6(2) & $-$3.2$\pm$0.5 & 1.0 & 1(2) & 30.64$\pm$1.4 & 13.99(2) & 1 & 0.458 & 0.201 & 0.014\\
J174656.41$-$282811.2 & earlyK & \nodata & 17$\pm$6 & 0.6(2) & $-$2.5$\pm$0.6 & 1.0 & 7(2) & 29.34$\pm$1.5 & 18.749(4) & 1 & 0.628 & 0.233 & 0.026\\
J174658.35$-$282617.3 & lateG? & \nodata & 13$\pm$5 & 0.6(2) & $-$4.4$\pm$0.5 & 0.9 & 2(2) & 29.25$\pm$1.5 & 13.70(7) & 1 & 0.582 & 0.389 & 0.017\\
J174659.34$-$282139.1 & F/G? & \nodata & 27$\pm$7 & 0.8(2) & $-$2.2$\pm$0.5 & 1.2 & 6(1) & 29.84$\pm$1.4 & 19.35(1) & 1 & 0.972 & 0.295 & 0.068\\
J174700.24$-$282236.0 & earlyM dMe?& \nodata & 14$\pm$5 & 0.5(2) & $-$1.8$\pm$0.6 & 0.9 & 9(2) & 29.30$\pm$1.5 & 20.05(1) & 1 & 0.906 & 0.237 & 0.062\\
J174700.72$-$283204.9 & midM dMe& \nodata & 444$\pm$22 & 0.2(2) & $-$1.4$\pm$0.4 & 0.3 & 10(1) & 29.61$\pm$1.9 & 16.86(1) & 1 & 0.518 & 0.390 & 0.020\\
J174706.77$-$283356.4 & midM & \nodata & 17$^{\star}\pm$8 & 0.1(1) & $-$2.7$\pm$0.4 & 0.3 & 10(1) & 27.94$\pm$2.1 & 17.02(3) & 1 & 2.231 & 1.546 & 0.438\\
J174708.21$-$282724.8 & G? & \nodata & 7$^{\star}\pm$4 & 0.8(2) & $-$2.5$\pm$0.5 & 1.2 & 6(1) & 29.42$\pm$1.4 & 19.78(1) & 1 & 0.564 & 0.423 & 0.017\\
J174709.30$-$282317.5 & F/G & \nodata & 10$\pm$5 & 0.8(1) & $-$2.9$\pm$0.5 & 1.2 & 5(1) & 29.68$\pm$1.4 & 18.13(1) & 1 & 0.905 & 0.199 & 0.088\\
J174710.66$-$282828.4 & early/midG & \nodata & 29$\pm$7 & 0.5(2) & $-$2.6$\pm$0.5 & 0.8 & 6(2) & 29.39$\pm$1.5 & 17.18(1) & 1 & 0.483 & 0.197 & 0.026\\
J174711.42$-$282533.9 & F/G? & \nodata & 8$^{\star}\pm$4 & 1.0(1) & $-$3.1$\pm$0.6 & 1.4 & 5(1) & 29.26$\pm$1.4 & 19.86(1) & 1 & 0.575 & 0.219 & 0.023\\
J174712.39$-$282836.9 & G? & \nodata & 23$\pm$6 & 0.9(2) & $-$2.5$\pm$0.5 & 1.3 & 5(2) & 30.00$\pm$1.3 & 18.94(1) & 1 & 0.538 & 0.287 & 0.022\\
J174713.11$-$282906.4 & K/M & \nodata & 7$\pm$4 & 0.4(1) & $-$2.1$\pm$0.4 & 0.7 & 9(1) & 28.81$\pm$1.6 & 19.46(1) & 1 & 0.733 & 0.132 & 0.035\\
J174713.29$-$282037.4 & G/K & \nodata & 53$\pm$9 & 0.8(2) & $-$2.2$\pm$0.5 & 1.2 & 5(1) & 30.26$\pm$1.3 & 18.37(1) & 1 & 0.812 & 0.115 & 0.026\\
J174714.12$-$282528.3 & midK & \nodata & 19$\pm$6 & 0.1(2) & $-$2.2$\pm$0.4 & 0.3 & 11(1) & 28.01$\pm$2.1 & 17.91(1) & 1 & 0.504 & 0.250 & 0.012\\
J174714.27$-$282108.4 & K/M dMe& \nodata & 31$\pm$8 & 0.4(2) & $-$2.1$\pm$0.5 & 0.6 & 8(2) & 29.11$\pm$1.6 & 18.12(1) & 1 & 0.977 & 0.101 & 0.075\\
J174714.77$-$282811.4 & ? & \nodata & 34$\pm$7 & 0.9(2) & $-$7.2$\pm$2.8 & 1.3 & $-$12(12) & 30.87$\pm$1.3 & 17.79(2) & 1 & 0.485 & 0.115 & 0.016\\
J174715.50$-$283155.7 & earlyM dMe& \nodata & 18$\pm$7 & 0.4(3) & $-$2.2$\pm$0.6 & 0.6 & 9(2) & 28.89$\pm$1.6 & 18.49(1) & 1 & 1.137 & 0.583 & 0.185\\
J174715.82$-$282643.1 & G/K & \nodata & 11$\pm$5 & 0.6(1) & $-$2.7$\pm$0.4 & 1.0 & 6.1(9) & 29.27$\pm$1.5 & 18.49(1) & 1 & 0.570 & 0.316 & 0.036\\
J174717.98$-$282729.3 & midM dMe& \nodata & 13$\pm$5 & 0.5(3) & $-$1.8$\pm$0.6 & 0.8 & 10(2) & 28.97$\pm$1.6 & 20.13(2) & 1 & 0.666 & 0.233 & 0.026\\
J174718.27$-$282753.3 & earlyG & \nodata & 16$\pm$6 & 0.4(1) & \nodata & 0.7 & 10(1) & \nodata & 19.82(2) & 1 & 0.622 & 0.611 & 0.023\\
J174719.59$-$282204.9 & lateG? & \nodata & 4$^{\star}\pm$5 & 0.9(2) & \nodata & 1.3 & 6(1) & \nodata & 20.05(1) & 3 & 4.128 & 3.930 & 1.874\\
J174720.10$-$282437.8 & ? & \nodata & 10$\pm$5 & 2.4(2) & $-$7.0$\pm$6.1 & 2.5 & \nodata & 31.60$\pm$1.8 & 20.08(3) & 1 & 0.941 & 0.437 & 0.077\\
J174720.16$-$282823.6 & midM dMe& \nodata & 11$\pm$5 & 0.1(1) & $-$1.3$\pm$0.4 & 0.3 & 15.0(8) & 27.70$\pm$2.1 & 20.99(2) & 1 & 0.867 & 0.207 & 0.068\\
J174721.24$-$281928.5 & ? & \nodata & 24$\pm$8 & $>$10.0$\dagger$ & \nodata & $>$6.0 & \nodata & \nodata & 20.75(2) & 1 & 2.477 & 1.219 & 0.388\\
J174722.25$-$282530.0 & lateF & \nodata & 12$\pm$5 & 0.2(2) & $-$4.1$\pm$0.5 & 0.4 & 6(2) & 28.05$\pm$1.9 & 14.21(1) & 1 & 0.821 & 0.141 & 0.052\\
J174723.15$-$282747.6 & ? & \nodata & 12$\pm$5 & $<$3.0$\dagger$ & $> -$5.1 & $<$3.0 & \nodata & $<$30.0 & 22.78(6) & 1 & 1.662 & 0.841 & 0.937\\
J174723.36$-$282534.0 & ? & \nodata & 62$\pm$9 & 2.3(2) & \nodata & 2.5 & \nodata & \nodata & 20.42(2) & 1 & 0.533 & 0.393 & 0.023\\
J174725.59$-$282930.8 & lateA & \nodata & 7$^{\star}\pm$5 & 0.6(2) & $-$3.7$\pm$0.5 & 1.1 & 4(1) & 29.16$\pm$1.4 & 16.32(1) & 1 & 1.436 & 0.318 & 0.238\\
J174726.89$-$283309.1 & midG & \nodata & 27$\pm$9 & 0.2(1) & $-$2.8$\pm$0.4 & 0.3 & 10.0(9) & 27.87$\pm$2.0 & 17.28(2) & 1 & 2.506 & 1.621 & 0.464\\
J174727.64$-$282646.5 & early/midG & \nodata & 34$\pm$7 & 0.5(2) & $-$2.7$\pm$0.4 & 0.8 & 6(1) & 29.46$\pm$1.5 & 17.05(1) & 1 & 0.680 & 0.253 & 0.025\\
J174731.11$-$282513.0 & mid/lateF & \nodata & 18$\pm$7 & 0.5(2) & $-$2.9$\pm$0.5 & 0.9 & 7(1) & 28.93$\pm$1.5 & 18.15(1) & 1 & 3.011 & 0.390 & 1.673\\
J174732.09$-$282033.8 & G/K? & $tt$ & 53$\pm$10 & 0.5(2) & $-$2.7$\pm$0.5 & 0.8 & 6(2) & 29.71$\pm$1.5 & 16.15(1) & 1 & 1.559 & 0.884 & 0.149\\
J174733.60$-$281840.6 & lateF? & \nodata & 48$\pm$11 & 0.4(1) & $-$4.0$\pm$0.4 & 0.8 & 2(1) & 29.65$\pm$1.5 & 12.98(1) & 1 & 1.994 & 1.531 & 0.123\\
J174734.60$-$283216.6 & K/M dMe& \nodata & 95$\pm$12 & 0.3(1) & $-$2.2$\pm$0.3 & 0.5 & 7.7(9) & 29.35$\pm$1.7 & 16.28(1) & 1 & 1.045 & 0.409 & 0.032\\
J174735.64$-$282011.4 & lateG & \nodata & 120$\pm$14 & 0.6(2) & $-$2.7$\pm$0.5 & 1.0 & 4(1) & 30.25$\pm$1.5 & 15.95(1) & 1 & 1.091 & 0.587 & 0.104\\
J174736.39$-$283214.0 & F/G & \nodata & 116$\pm$14 & 0.3(1) & $-$2.4$\pm$0.3 & 0.4 & 6.9(9) & 29.37$\pm$1.7 & 15.55(1) & 1 & 1.015 & 0.799 & 0.050\\
J174737.58$-$282228.7 & midK & \nodata & 20$\pm$8 & 0.2(2) & $-$2.3$\pm$0.5 & 0.3 & 10(2) & 28.28$\pm$2.0 & 17.57(1) & 1 & 1.699 & 0.575 & 0.183\\
J174737.90$-$282040.6 & ? & \nodata & 43$\pm$11 & $<$2.0$\dagger$ & $>$2.5 & $<$2.3 & $>$0.86 & $<$31.0 & 21.59(2) & 1 & 1.689 & 1.170 & 0.174\\
J174746.14$-$282350.5 & earlyG & \nodata & 20$^{\star}\pm$10 & 0.8(1) & \nodata & 1.2 & 5(1) & \nodata & 18.44(1) & 1 & 2.483 & 1.400 & 0.826\\
\sidehead{GalCA}
J174614.20$-$285136.4 & lateG & \nodata & 127$\pm$13 & 0.7(1) & $-$2.0$\pm$0.3 & 1.1 & 4.4(9) & 30.70$\pm$1.4 & 16.97(1) & 1 & 0.407 & 0.258 & 0.010\\
J174611.12$-$284933.9 & G? & \nodata & 10$^{\star}\pm$5 & 0.5(2) & $-$2.2$\pm$0.5 & 0.8 & 7(2) & 29.40$\pm$1.5 & 18.45(1) & 1 & 1.161 & 0.707 & 0.081\\
J174618.87$-$285145.9 & late/midF & \nodata & 12$\pm$5 & 0.1(1) & $-$3.6$\pm$0.4 & 0.2 & 8(1) & 27.62$\pm$2.3 & 14.10(1) & 1 & 0.489 & 0.137 & 0.015\\
J174626.49$-$285411.1 & early/midA & \nodata & 3$^{\star}\pm$4 & 0.3(2) & $-$3.8$\pm$0.5 & 1.1 & 4(1) & 29.09$\pm$0.9 & 15.04(1) & 1 & 0.685 & 0.027 & 0.032\\
J174629.36$-$285100.8 & G?(by) & $by$ & 12$\pm$5 & 0.5(2) & $-$1.6$\pm$0.5 & 0.8 & 9(2) & 29.54$\pm$1.5 & 19.57(1) & 1 & 0.594 & 0.097 & 0.024\\
J174630.19$-$285442.1 & midM & \nodata & 23$\pm$6 & 0.2(1) & $-$1.6$\pm$0.4 & 0.3 & 12.2(9) & 28.41$\pm$2.0 & 18.82(1) & 1 & 0.636 & 0.096 & 0.026\\
J174637.44$-$285108.7 & midM dMe& \nodata & 13$\pm$5 & 0.00(1) & $-$1.6$\pm$0.3 & 0.1 & 16.6(5) & 26.71$\pm$0.2 & 18.61(1) & 1 & 0.895 & 0.416 & 0.057\\
J174638.02$-$285326.2 & CV-A & $cv$ & 438$\pm$22 & 1.4(3)$\dagger$ & 1.2$\pm$0.6 & 1.8 & 3(2) & 31.87$\pm$1.2 & 20.23(2) & 1 & 0.447 & 0.317 & 0.013\\
J174656.89$-$285233.9 & CV-C & $cv$ & 30$\pm$12 & $>$1.4$\dagger$ & $< -$2.3 & $>$1.7 & $<$3.99 & $>$30.2 & 21.30(2) & 1 & 2.36 & 2.03 & 0.401\\
\sidehead{SgrA}
J174511.51$-$290236.8 & K/M dMe& \nodata & 83$\pm$18 & 0.8(5) & $-$2.5$\pm$0.9 & 2.3 & 5(4) & 30.21$\pm$0.7 & 19.23(1) & 1 & 0.899 & 0.868 & 0.045\\
J174516.18$-$290315.8 & ? & \nodata & 728$\pm$32 & 6(1) & $-$7.6$\pm$2.1 & 11.5 & \nodata & 33.64$\pm$0.7 & 22.31(5) & 1 & 0.497 & 0.221 & 0.015\\
J174517.75$-$290139.3 & lateG? & \nodata & 212$\pm$19 & 0.2(3) & $-$2.4$\pm$0.6 & 1.3 & 6(2) & 29.75$\pm$0.6 & 17.04(1) & 1 & 0.525 & 0.111 & 0.016\\
J174520.62$-$290152.1 & midM dMe& \nodata & 1071$\pm$35 & 0.3(3) & $-$1.3$\pm$0.6 & 1.4 & 7(2) & 30.47$\pm$0.7 & 18.22(1) & 1 & 0.429 & 0.209 & 0.011\\
J174522.76$-$290018.5 & ? & \nodata & 107$\pm$14 & $<$0.001$\dagger$ & $> -$2.3 & $<$0.1 & $>$13.7 & $<$26.9 & 17.83(1) & 1 & 0.505 & 0.151 & 0.013\\
J174528.08$-$285726.5 & midM dMe& \nodata & 136$\pm$16 & 0.00(6) & $-$2.0$\pm$0.3 & 0.1 & 15.8(7) & 26.52$\pm$0.0 & 17.97(1) & 1 & 0.475 & 0.034 & 0.011\\
J174530.63$-$290441.0 & F/G & \nodata & 127$\pm$16 & 0.4(3) & $-$2.7$\pm$0.6 & 1.4 & 5(2) & 29.69$\pm$0.7 & 17.17(1) & 1 & 0.556 & 0.234 & 0.019\\
J174537.99$-$290134.5 & A/F? & \nodata & 290$\pm$22 & 3(1) & $-$5.2$\pm$1.7 & 4.5 & $-$7(7) & 32.08$\pm$0.8 & 18.45(1) & 1 & 0.383 & 0.105 & 0.009\\
J174538.28$-$285602.7 & midM dMe& \nodata & 202$\pm$18 & 0.1(3) & $-$1.5$\pm$0.5 & 0.9 & 11(2) & 29.17$\pm$0.5 & 19.40(1) & 1 & 0.476 & 0.289 & 0.012\\
J174540.88$-$290328.9 & midM dMe?& \nodata & 36$\pm$10 & 0.6(4) & $-$2.2$\pm$0.7 & 1.6 & 7(3) & 29.40$\pm$0.9 & 20.02(1) & 1 & 0.518 & 0.195 & 0.017\\
J174545.65$-$285621.7 & midM dMe& \nodata & 70$\pm$14 & 0.2(3) & $-$2.1$\pm$0.5 & 1.1 & 9(2) & 29.12$\pm$0.6 & 18.62(1) & 1 & 0.519 & 0.114 & 0.016\\
J174548.61$-$290522.4 & lateG? & \nodata & 273$\pm$22 & 0.4(2) & $-$2.0$\pm$0.5 & 1.4 & 6(2) & 30.21$\pm$0.7 & 17.74(1) & 1 & 0.516 & 0.014 & 0.019\\
J174549.18$-$285557.4 & F/G & \nodata & 89$\pm$13 & 0.3(3) & \nodata & 1.4 & 6(1) & \nodata & 17.28(1) & 1 & 0.980 & 0.598 & 0.059\\
J174552.92$-$290358.8 & lateG? & \nodata & 831$\pm$33 & 0.3(2) & $-$2.3$\pm$0.5 & 1.4 & 4(2) & 30.39$\pm$0.7 & 16.12(1) & 1 & 0.431 & 0.096 & 0.015\\
J174558.23$-$285644.3 & early/midK & \nodata & 185$\pm$22 & 0.4(1) & $-$2.3$\pm$0.4 & 1.5 & 6(1) & 29.85$\pm$0.7 & 17.79(1) & 1 & 0.536 & 0.114 & 0.019\\
J174559.56$-$285435.8 & G/K & \nodata & 501$\pm$34 & 1.5(2) & $-$3.5$\pm$0.6 & 3.2 & \nodata & 30.04$\pm$0.9 & 20.16(4) & 1 & 0.575 & 0.506 & 0.017\\
J174607.52$-$285951.3 & CV-B & $cv$ & 3539$\pm$63 & 0.7(4)$\dagger$ & $-$0.1$\pm$0.9 & 1.9 & 8(3) & 31.25$\pm$0.9 & 21.75(3) & 1 & 0.430 & 0.007 & 0.013\\
\enddata
\tablenotetext{a}{\scriptsize{ChaMPlane IDs have prefix ChOPS...}}
\tablenotetext{b}{\scriptsize{See Table 4 for explanation of spectral type classification.}}
\tablenotetext{c}{\scriptsize{Additional classification of source type. q: qLMXB candidate, by: BY Dra candidate, rs: RS CVn candidate, tt: T Tauri, cv: CV.}}
\tablenotetext{d}{\scriptsize{Background subtracted net counts in the Bx band (0.3--8.0keV). $^{\star}$This source has SNR$<$3.}}
\tablenotetext{e}{\scriptsize{Hydrogen column density $N_H$ in units of 10$^{22}$ cm$^{-2}$ estimated from spectral fit. True value may be higher by $\sim$0.2, see text. $^{\dagger}$ Objects with `?' spectral type and CVs have $N_H$ estimated from QCCD analysis.}}
\tablenotetext{f}{\scriptsize{$\log$ of the ratio of unabsorbed flux in S$_C$band 0.5--2.0keV to unreddened optical R band flux (ergs cm$^{-2}$ s$^{-1}$ (1000$\textrm{\AA}$)$^{-1}$)}}
\tablenotetext{g}{\scriptsize{The distance in kpc. Error $\approx$60\%.}}
\tablenotetext{h}{\scriptsize{$\log$ of X-ray luminosity in S$_C$ band (ergs s$^{-1}$)}}
\tablenotetext{i}{\scriptsize{NM---number of optical matches found within 95\% confidence (2$\sigma$) error circle of this X-ray source.}}
\tablenotetext{j}{\scriptsize{`Srch': the combined 95\% X-ray and optical position error circle (arcsec) used to search for optical matches. `Offset': the positional offset in arcsec between X-ray and optical position. `P$_{Rn}$': random match probability given error circle size and measured local optical projected surface density.}}
\end{deluxetable}

\subsection{Quantile Analysis}
For stars with low signal-to-noise ratio optical spectra we assign in
Table 5 a non-classification to the object `?'. In order to place
constraints on $N_H$ for these sources, and for the CVs, whose optical
spectra are not well-modelled by single-temperature blackbodies, we
utilize X-ray quantile analysis. Objects for which this has been
carried out are marked with a $`\dagger'$ in Table 5. Quantile
analysis was first presented in \citet{hong04}. It allows us to derive
X-ray spectral information despite low source counts. It involves
placing sources on an X-ray color-color diagram by the median and
quartile energy fractions of their source counts. As defined by H04,
any general quantile Q$_x$ is calculated as:
\begin{equation}
Q_x = \frac{E_{x\%} - E_{lo}}{E_{up} - E_{lo}}
\end{equation}
where E$_{x\%}$ is the energy below which the net counts is $x\%$ of
the total number of counts between E$_{lo}$ and E$_{up}$; we select
E$_{lo}$=0.3 and E$_{up}$=8.0 keV (B$_X$ band) for our analysis. We
plot the ratio Q$_{25}$/Q$_{75}$ against $\log_{10}$[m/(1$-$m)]
(m(${\equiv}$Q$_{50}$) is the median). For a given spectral model, we
overlay a grid of column density $N_H$ and model parameter---we
interpolate to find a value for $N_H$. The grid shape is dependent on
the $\it{Chandra}$ ACIS response function for the ObsID
considered.

Figure 4 shows QCCD plots for ObsID 53392, which includes CV-B
(detected with $\sim$3500 counts in this ObsID) and ObsID 945, which
includes all three CVs we have detected in the {\it Chandra} fields
considered in this paper. We use a thermal bremsstrahlung model to
construct the grids shown in this case.

\subsection{Extinction Model for the Galaxy}
We require a method of deriving luminosity for our sources and so need
a way of estimating their distances. Recently, \citet{drimmel01}
(hereafter D01) constructed a three-component model for the dust
distribution in the Galaxy. This is used via a FORTRAN code
\citep[][hereafter: D03]{drimmel03} to derive the extinction, $A_V$ as
a function of distance from the Sun over the whole sky. The spatial
resolution on which values of $A_V$ can be measured is $\sim$20
arcmin. Figure 5 shows the run of $A_V$ with distance for each field
in this paper. After fitting the optical spectra or using quantile
analysis to derive extinction values $A_V$, we utilize the model of
D01 to derive distances to all sources in this paper. The values
listed in Table 5 are {\it not} corrected for the overestimate in
$A_V$. As a check on the results, the spectral fit scaling factor
$R^2$/$d^2$ allows us to derive a radius which we compare with
expected values for stars of given spectral type and $M_V$
\citep[e.g.][]{Allen2000}.

The code has a `rescaling' option to account for small-scale
variations and clumping in the dust density that are smoothed over by
the model. This is based on a factor dependent on the residuals
between the COBE observed flux at 240$\mu$m and the model prediction
for the same. It was found that for all fields, using the rescaling
option in the D03 code worked well in producing reasonable stellar
radii within 50\% of tabulated values \citep[e.g.][]{Allen2000}, with
the exception of G347b, where radii were systematically a factor of
four to five times higher. 

For G347b we derive $A_V(d)$ using extracted emission spectra of
Galactic molecular CO \citep[from the survey of][]{dame01} and 21cm HI
diffuse gas \citep[data taken from the Southern Galactic Plane
Survey,][]{taylor03} to derive the column density of molecular and
atomic hydrogen respectively, thus $N_H$ via $N_H$ = $N_{HI}$ +
$N_{H_2}$ and so $A_V$. The HI data has a resolution of $\sim$1$'$,
but for this analysis was smoothed to 3$'$ resolution with 2.7$'$
spacing. Emission from CO was assigned a distance based on
line-of-sight velocity, splitting the near/far ambiguity based on the
latitude of observation, and using the Galactic rotation curve of
\citet{brand93}. We assumed a FWHM layer thickness for HI of 220pc
\citep{dickey90}, and for CO of 120pc \citep{dame01}. Emission beyond
the terminal velocity cutoff was redistributed in a Gaussian below the
cutoff with the Gaussian dispersion equal to the cloud-cloud velocity
dispersion: $\sigma$(CO) = 4 km s$^{-1}$, $\sigma$(HI) = 8 km
s$^{-1}$. We assume an atomic hydrogen spin temperature of 140K. We
performed this calculation along two lines of sight---in Galactic
coordinates these are at (l, b): 347.375, $-$0.75 and 347.375,
$-$0.875 (positions 1 and 2 respectively). Figure 5, right-hand panel
shows the resultant $A_V$ versus distance plots for field G347b,
labelled position 1 and position 2. We were unable to perform the same
derivation for the other fields as the three Galactic Center fields
(SgrA$\star$, SgrB2 and GalCA) are too close to the Galactic plane to
easily assign distances to molecular emission, and for field J1655 the
HI data available was not at high enough resolution to accurately run
the calculation.

\subsection{Calculating X-ray Fluxes}
We calculate unabsorbed X-ray fluxes for all sources from their net
count rates using
$\emph{sherpa}$\footnote{http://cxc.harvard.edu/sherpa/threads/index/html}.
For simplicity, we assume that the X-ray radiation produced by the
majority of objects in our sample will be emission from a hot
(T$>$10$^6$ K, kT$>$0.1 keV) coronal plasma. To calculate fluxes in
the hard (H$_C$) and soft (S$_C$) bands for each star in our sample we
thus adopt a simple single-temperature MEKAL model\footnote{Model:
  xsmekal in sherpa} \citep[bremsstrahlung emission of an optically
thin, thermal plasma with metal absorption and emission lines;
see][]{mewe85} at 1keV and use the $N_H$ value listed in Table
5. $\S$3.4 below discusses the uncertainty introduced by our choice of
spectral model on our results for X-ray flux. We derive X-ray
luminosities via $L_x$ = 4$\pi$ $d^2$ $F_x$ ergs s$^{-1}$. The X-ray
to optical flux ratio is calculated via:
\begin{equation}
\log(F_x/F_R) = \log(F_x) + 0.4R + 5.765
\end{equation}
and
\begin{equation}
\log(F_x/F_V) = \log(F_x) + 0.4V + 5.426
\end{equation}
where we have assumed a square optical filter transmission function of
width 1000$\textrm{\AA}$, centered on the filter's quoted central
wavelength, with an underlying A0 stellar spectrum to calculate the
constants. Note that the overestimate in $N_H$ from $\S$ 3 carried
through results in $\Delta \log(F_x/F_V)$ = +0.3 -- +0.5 and $\Delta
\log(L_x) \sim -$0.3. The exact correction depends on the object's
initial $N_H$. This is {\it not} incorporated in the results given in
Table 5.

\subsection{Error Analysis}
The primary source of error in our analysis is the calculation of
$E(B-V)$---this has three components. Firstly, there is the
uncertainty in assigned T$_{eff}$ values ($\pm \sim$500K). Secondly,
from systematic errors in the flux calibrated spectra (from the
extraction and calibration processes). The flux calibration error
ranges from $\sim$10 to 60\%---estimated by comparing repeat
observations of stars observed on multiple nights. This was the case
for moderate and also high S/N spectra. Variations were evident in
spectra both within a night as well as from night to night. Thirdly,
errors introduced by our spectral fitting code. We estimate this by
selecting 17 standard star spectra (types M5 V through A7 V) from the
catalog of \citet{jacoby84}, applying a range of fake values of
interstellar reddening ($E(B-V)$ = 0.3--4.0) with the
$\emph{fm\_unred}$ command in
IDL\footnote{http://idlastro.gsfc.nasa.gov/} and then attempting to
retrieve this fake reddening with our spectral fit code. The
$`$error$'$ on the returned value $\Delta E(B-V) \approx 50\%$ at
$E(B-V) \approx$ 0.3, ranging down to $\approx 6\%$ at $E(B-V)
\approx$ 1.3 (independent of spectral type). We linearly interpolate
this trend to calculate the uncertainty produced by the fitting
process at any $E(B-V)$. The quoted error in column density $N_H$ in
Table 5 in column 5 combines all these three sources.

Errors quoted in Table 5 for $\log(F_x/F_R)$ and $\log(L_x)$
incorporate uncertainty in $N_H$, in the X-ray count rate and
photometric error in the $R$ magnitude (a relatively small
contribution). The choice of X-ray spectral model (1.0keV MEKAL) also
contributes to our uncertainty. Using a power law model with spectral
index $\Gamma$=1.7 or a MEKAL (kT=2.0 keV) reduces the S$_C$ band flux
by between 10\% (at $N_H \approx $ 0.2$\times$10$^{22}$ cm$^{-2}$) and
50\% (at $N_H \approx $ 2.0$\times$10$^{22}$ cm$^{-2}$). A MEKAL model
with kT=0.5 keV creates a similar {\it increase} in the S$_C$ band
flux. Thus the errors quoted in $\log(F_x/F_R)$ and $\log(L_x)$ should
be considered as lower limits.

Using the model of \citet{drimmel01} introduces error in our distance
calculation. Results for $d$ derived from this model for Galactic
longitudes $\mid$l$\vert$$<$ 20$\degr$ are likely to suffer from
significant systematic uncertainty as true structure in the absorbing
interstellar medium is significant in this part of the Galaxy and is
poorly modeled in the code. The fact that fields SgrB2 and GalCA
($\sim$0.5$\degr$ apart of the sky) are indistinguishable on this
figure and the curve for field SgrA$\star$ (which is only $\sim$12$'$
away) is significantly different is almost certainly a resultant
artifact, not a real trait of these two three fields. The combined
error from spectral fit and dust model (estimated by comparing
variation in $A_V$ across the {\it Chandra} field of view) in the
distances we derive is typically $\lesssim$60\%. We adopt 60\% as the
error on the distances that we quote in Table 5. This carries through
to error in X-ray luminosity and optical absolute magnitudes.

\section{Results}
Table 5 presents our spectral results. Optical photometry comes from
data from our 2000 and 2003 CTIO Mosaic runs
\citep[see][]{zhao05}. For each source in Table 5 we list the
ChaMPlane IAU optical source ID (column [1]). Readers desiring to find
detailed X-ray counterpart properties can readily search our online
X-ray
database\footnote{http://hea-www.harvard.edu/ChaMPlane/database$\_$xray.html}
using the optical source ID for coordinates. We then give the source
properties: spectral classification (column [2]), net X-ray source
counts in the B$_X$ band (column [3]), the hydrogen column density
$N_H$ as derived from our fitting technique or quantile analysis
(column [4]), the unabsorbed X-ray to $R$-band flux ratio in the S$_C$
band, adopting a 1.0 keV MEKAL X-ray spectral model (column [5], see
$\S$ 2 for band definitions). Although subsequent analysis is
performed on a derivation of $\log (F_x/F_V)$ for easier comparison
with the literature, the actual numbers do not differ significantly
from those listed for $\log (F_x/F_R)$. In addition, since $\sim$10\%
of our objects are undetected in $V$, we can give a value for more
sources by presenting this information instead. The distance and the
derived absolute visual magnitude M$_V$ are in columns [6] and [7]. We
then give X-ray luminosity, followed by the optical $R$ magnitude and
the number of optical sources found to match the same X-ray position
(columns [8], [9] and [10]). Column [11] gives the X-ray 2 $\sigma$
search radius size in arcseconds and column [12] gives the offset of
the optical position in arcsec from the center of the X-ray error
circle. Column [13] gives the expected number of optical sources that
should fall in an error circle of this size by chance, given the
observed surface density of stars within 1 arcmin of the X-ray
position on the Mosaic image. For almost all objects in Table 5 one
optical source matches the X-ray position, however, in the J1655
field, ChOPSJ165407.44$-$394542.7 and J165350.37$-$394621.9, in the
G347b field ChOPSJ171524.21$-$395950.9 and in the SgrB2 field
ChOPSJ174719.59$-$282204.9, J174633.29$-$282512.7 and
J174634.65$-$282721.1 are all one of multiple matches to single X-ray
sources. In the case of ChOPSJ165350.37$-$394621.9 in J1655 we have a
spectrum of the other match---we select the M-type star as the more
likely X-ray emitting candidate over the G-type alternate. The
`Classification' column follows the scheme: mid G represents G4--G6,
late G: G6--G8, F/G: F8--G2 and so on. A question mark placed next to
a classification indicates that the uncertainty in spectral type is
greater so mid G?  means G3--G7 and F/G? means F7--G3.  A star
classified as G? means G0--G9 (and equivalently for other types).

\subsection{Cataclysmic Variables}
The full list of matches between our X-ray and optical source lists
having $H\alpha - R < -$0.3, $V <$ 23 and an optical signal to noise
ratio $>$ 1.4 is given in Table 6 below\footnote{CV-C is included for
  its broad H$\alpha$ emission line.}. We restrict $H\alpha - R$ based
on the work of \citet{szkody04} who find that only 17\% of their
sample of CVs from the Sloan Survey have an $H\alpha$ equivalent width
below 28$\textrm{\AA}$ in emission (i.e. have $H\alpha - R >
-0.3$). We can use spectra to rule out 4 of these as dMe, normal or T
Tauri stars. The objects: ChOPSJ174559.18$-$290418.9 in field
SgrA$\star$ and ChOPSJ165335.32$-$393715.9 in field J1655 had no
optical spectra taken in any ChaMPlane observing run and so remain
uncertain CV candidates.

\begin{deluxetable}{lcccc}
\tablewidth{0pt}
\tablecolumns{5}
\tabletypesize{\small}
\tablenum{6}
\tablecaption{$H\alpha -R < -$0.3 Objects} 
\tablehead{ \colhead{ChOPS ID} & \colhead{$H\alpha -R$} & \colhead{$R$} & \colhead{Spectrum} & \colhead{$\log(F_x/F_R)$}}
\startdata
\sidehead{J1655}
335.32-393715.9 & $-$0.46(6) & 21.5 & None & $<-$0.99$^{\star}$\\
422.01-395205.0 & $-$0.39(2) & 19.6 & dMe & $-$1.5$\pm$0.5\\
\sidehead{GalCA}
638.02-285326.2 & $-$0.65(2)$^{\dagger}$ & 20.2 & CV(A) & $1.2\pm0.6$\\
656.89-285233.9 & $-$0.19(4)$^{\ddagger}$ & 21.3 & CV(C) & $< -$2.3\\
\sidehead{SgrB2}
732.09-282033.8 & $-$0.61(1) & 16.2 & T Tauri & $-2.7\pm0.5$\\
708.21-282724.8 & $-$0.38(1) & 19.8 & G? & $-2.5\pm0.5$\\
\sidehead{SgrA}
511.51-290236.8 & $-$0.31(1) & 19.2 & dMe & $-2.5\pm0.9$\\
559.18-290418.9 & $-$0.6(1) & 22.7 & None & $-1.5\pm0.5$ \\
607.52-285951.3 & $-$0.61(4) & 21.8 & CV(B) & $-0.1\pm0.9$\\
\enddata
\tablecomments{Numbers in parentheses with $H\alpha - R$ value represent error in last quoted digit. $^{\star}$Undetected in S$_C$ band---estimated upper limit to flux ratio. $^{\dagger}$This value of $H\alpha -R$ was recorded in 2000. In 2003 it had dropped to $-$0.182. $^{\ddagger}$CV-C is included here despite having $H\alpha - R>-$0.3.}
\end{deluxetable}

Qualitative analysis of the remaining spectra reveals two clear CV
candidates in the GalCA field from the LDSS2 spectral sample (see
Table 5, sources ChOPSJ174638.02$-$285326.2 and
...656.89-285233.9). Hereafter we refer to these as CV-A and CV-C
respectively (see Figures 6a and 6c). Another source,
ChOPSJ174607.52-285951.3 (hereafter, CV-B), was imaged with LDSS2 in
the SgrA$\star$ field, but the spectrum suffered from being on the
edge of the slit and was not possible to extract. It was later
re-observed by the IMACS instrument at Magellan---its IMACS spectrum
is shown in Figure 6b. All three spectra show broad $H\alpha$ in
emission. CV-A and CV-B also show emission lines of HeI.

CV-B was previously detected by ROSAT Position Sensitive Proportional
Counter (PSPC) observations of the GC regions \citep{sidoli01}, as
their source `65,' but without further identification possible at the
time. It was found to have a count rate of 1.9$\pm$0.3 cts ks$^{-1}$
in the ROSAT band 0.1--2.4keV. Using the online PIMMS tool with an
assumed thermal bremsstrahlung spectrum with kT = 7.3 keV and $N_H$ =
5$\times$10$^{21}$ cm$^{-2}$ from our QCCD analysis (Figure 4), this
converts to a {\it Chandra} hard-band (2--8 keV) flux of 1.0$\pm$0.2
$\times$ 10$^{-13}$ erg s$^{-1}$ cm$^{-2}$. From our {\it Chandra}
data we estimate 0.64$\pm$0.15 $\times$ 10$^{-13}$ erg s$^{-1}$
cm$^{-2}$, possibly indicative of some slight variability.

We use the QCCD results of Figure 4 to estimate by eye a plausible
initial spectral model (CV-A and CV-B were detected in ObsID 945 with
$\sim$300 counts, CV-C has only $\sim$26 counts). If we assume that a
bremsstrahlung spectrum is representative of their X-ray emission, we
can thus estimate their spectral properties (see Table 7). We present
XSPEC\footnote{http://xspec.gsfc.nasa.gov/docs/xanadu/xspec/index.html}
fits for the two X-ray bright CVs (CV-A and CV-B) in Figure 7
below. We use a bremsstrahlung emission plus photon absorption model
for the fit to each spectrum. CV-C has too few counts to provide
adequate signal to noise for spectral fitting. The results we derive
are shown in Table 7 below. For CV-A and CV-B, the quantile-derived
parameters are in good agreement with the estimates from the XSPEC
fits. For CV-C, we can only place weak constraints on $N_H$ and kT,
but the error bars are consistent with a typical CV spectrum of
bremsstrahlung at $\sim$2--8keV. This is suggestive that these CVs are
all dwarf novae (DN)---typically characterized by hot ($\sim$10$^8$K)
hard X-ray spectra in quiescence \citep{warner95}. We note however,
that the absolute magnitudes $M_V$ for CV-A and CV-C are on the bright
side for quiescent DN (3 and $<$3.99 respectively).

\begin{deluxetable}{clccccccc}
\tablecolumns{11}
\tabletypesize{\tiny}
\tablenum{7}
\tablewidth{0pt} 
\tablecaption{CVs With LDSS2 Spectra} 
\tablehead{ \colhead{CV} & \colhead{ChOPS ID\tablenotemark{a}} & \colhead{H$\alpha-R$} & \colhead{FWHM\tablenotemark{b}} & \colhead{EW\tablenotemark{c}} & \multicolumn{2}{c}{XSPEC} & \multicolumn{2}{c}{QCCD}\\
 & & (mag)& H$\alpha$($\textrm{\AA}$) & H$\alpha$($\textrm{\AA}$) & N$_H$\tablenotemark{e} & kT(keV)& N$_H$\tablenotemark{e} & kT(keV)}
\startdata
CV-A & 638.02-285326.2 & $-$0.66(3)\tablenotemark{c} & 26$\pm$3 & $-$48$\pm$5 & 1.0$\pm$0.2 & 8.9$\pm$4.6 & 1.4$\pm$0.3 & 9$\pm$3\\
CV-B & 607.52-285951.3 & $-$0.63(4) & 24$\pm$3 & $-$81$\pm$7 & 0.5$\pm$0.03 & 7.3$\pm$0.9 & 0.7$\pm$0.4 & 8$\pm$1\\
CV-C & 656.89-285233.9 & $-$0.20(4) & 31$\pm$3 & $-$60$\pm$10 & \nodata & \nodata & $>$1.4 & \nodata\\
\enddata
\tablenotetext{a}{Abbreviated IDs: prefix ChOPSJ174...}
\tablenotetext{b}{Full width at half maximum intensity of the H$\alpha$ line}
\tablenotetext{c}{Equivalent width of the H$\alpha$ line.}
\tablenotetext{d}{As of 2000. In 2003 H$\alpha-$R was measured at -0.194.}
\tablenotetext{e}{N$_H$ in units of 10$^{22}$cm$^{-2}$}
\end{deluxetable}

The final tally of likely CVs is thus: 1 (possibly 2) in SgrA$\star$,
2 in GalCA and possibly 1 in the J1655 field. This latter object lies
outside the main ACIS-S chip (S3) and its neighboring S4 chip---the
large error circle size in which it is found means that even when
looking exclusively at objects with $H\alpha - R < -0.3$ in the field,
the probability it is a random match is high ($>$20\%). We consider
this a low probability CV candidate.

\subsection{Stellar X-ray Sources}
In order to compare our sample of stellar X-ray sources with other
surveys we construct their luminosity functions and assemble an X-ray
to optical flux ratio histogram.

Since our survey is flux limited, we correct for incompleteness with
the 1/$V_{max}$ method of \citet{schmidt68}. In this method, each
source contributes 1 over the maximum volume in which it could have
been detected in our survey to its bin in the cumulative luminosity
histogram. This maximum volume is calculated in the following way:
given one ACIS pixel in a {\it Chandra} observation, with a known
limiting (3$\sigma$) count rate for detection (and thus limiting flux,
given a spectral model) we can calculate a maximum distance d$_{max}$,
that an object of known luminosity could be placed and still be
detected. This defines a pyramidal volume given by the size of the
pixel on the sky. To calculate the full volume, we simply repeat this
process across the whole field of view, using the 3$\sigma$ count rate
limit at each point and sum the resulting volume elements together. To
speed the calculation up we re-bin each {\it Chandra} image by a
factor 80.

We divide the sources into intermediate (early F to late G) and late
(G/K to M) types, and plot the resultant, V$_{max}$-corrected
luminosity functions in Figure 8 below. We use luminosity calculated
in the ROSAT band (0.1--2.4keV) for ease of comparison with past
surveys, despite the mismatch in bandpass more suitable to our
reddened fields. We overplot comparison luminosity functions of main
sequence stars \citep[from the volume limited survey of][]{schmitt04},
and young main-sequence stars from the ROSAT surveys of the Hyades
\citep[age $<$ 1Gyr][]{stern95} and the Pleiades \citep[age
$\sim$10$^8$ yrs][]{micela96}. We convert their quoted luminosities
into the 0.1--2.4 keV band using the online PIMMS
tool\footnote{http://cxc.harvard.edu/toolkit/pimms.jsp}. Since in $\S$
3.1 we concluded that the spectral fit $N_H$ overestimates the true
value, for this comparison we generate luminosities with $N_H$ reduced
by 0.2$\times$10$^{22}$ cm$^{-2}$, and consequently {\it distances}
reduced appropriately given the D01 model.

In Figure 9 we plot a comparative histogram of the X-ray to optical
flux ratio ($V$-band flux), i.e. $\log(F_X/F_V)$. Again we utilize the
reduced $N_H$ values to generate X-ray rate-to-flux conversions to
calculate the X-ray fluxes, and de-redden the optical magnitudes. We
overplot the samples of \citet{schmitt04}, \citet{stern95} and
\citet{micela96}, scaled so that the total area under each curve is
the same for all histograms within a plot.

\subsection{Candidate Low Mass X-ray Binary}
A quiescent low-mass X-ray binary (qLMXB) system consisting of a black
hole or neutron star and a main sequence star will most likely show
strong H$\alpha$ in emission in its optical spectrum. Some CVs and
qLMXBs have sub-giant companions; these are more likely to be detected
in our reddened sample. Since the only objects showing H$\alpha$ in
emission in this sample are either CVs or T Tauri stars (and
identified as such in Table 5), we look to the example of GRO
J1655$−$40 \citep{zhang94,harmon95,bailyn95} a known qLMXB, soft X-ray
transient system, consisting of an F3--6IV sub-giant star secondary
and an accreting black hole primary. This system shows H$\alpha$ in
absorption in quiescence: the secondary star is luminous enough to
hide the emission line produced by the accretion disk. Our spectral
classification is not precise enough to ascertain the luminosity
classes of stars in the sample, so to find analogs to this system we
search for stars with spectral type earlier than K, with absolute
magnitude $M_V$ more than 2$\sigma$ higher than that expected for a
main sequence star of that type (see Table 4). We then look for stars
with $\log(F_X/F_R)$ (S$_C$ band) more than 2$\sigma$ greater than
that seen in sub-giant stars in the survey of H\"unsch et
al. (1998a). In Table 5 we flag the one candidate:
ChOPSJ165408.14$-$395636.1 that we find after this search. Figure 10a
presents the LDSS2 spectrum. This is only a very tentative
classification---it is also possible that this object is an active
binary of RS CVn type \citep{hall76}. We note that on the basis of its
spectrum, and X-ray properties in quiescence alone, GRO J1655$-$40
(see Figure 10b) could be mistaken for an RS CVn system. Further
variability analysis and detailed spectral followup is necessary to
rule out this object as a black hole or neutron star binary system.

\section{Discussion}
\subsection{Stellar Coronal Emission}
We have discovered a large sample of stellar coronal emission sources
in our survey fields. Recent studies, \citep[for
example,][]{taglia94,sciortino95} have shown that flux-limited X-ray
surveys of the Galaxy (within the disk) will naturally sample
preferentially from the youngest stellar populations present in the
Galactic disk, owing to the known decline of stellar X-ray emission
with age \citep{vaiana92}. Figures 7 and 8 show that the objects
discovered in our survey are somewhat elevated in their X-ray
emission, hence one explanation is that they are young. In this
picture, the luminosity functions in Figure 8 are possibly explained
by a composite of local (old) and younger population (Hyades and
Pleiades age) stars. However, the presence of very luminous objects
($\log (L_X) > $ 30) suggests an additional component, as does the
excess of objects at high $\log(F_X/F_V)$ in Figure 9.

The first possibility is that we are detecting an even younger
population, in other words a component of pre-main-sequence (pre-MS)
stars. Results from ROSAT and the COUP \citep{getman05} show that T
Tauri stars (weak-lined and classical) can have X-ray luminosities in
excess of 10$^{30}$ erg s$^{-1}$, and in some cases greater than
10$^{31}$ erg s$^{-1}$. Since our survey covers fields at low Galactic
latitude ($|b| <$ 3$\degr$), and since the scale height of stars also
increases with age \citep{wielen77}, this is a likely source of X-ray
active objects. To further test the likely pre-MS content of our
survey, we examined the near-infrared (near-IR) colors of ChaMPlane
objects in this paper \citep[using online data from the 2MASS
survey,][]{skrut06}. In Figure 11 we present a plot of $J-H$ versus
$H-K_S$ colors obtained in this way. We found 92 2MASS matches to our
136-object source-list. Overlaid on the plot are the locus of
Main-Sequence stars and giants from \citet{bessell88}, and a reddening
vector created using the near-IR extinction relation of
\citet{nishi06}. Objects below the line are candidate infrared excess
(T Tauri) young stars \citep[following the work of][]{lada92}---a
total of between 36 and 56, or 39--61\% of sources (the range is due
to the 2MASS photometric uncertainty). It is also possible for T Tauri
stars to be found above the line---this component is not possible to
establish without high resolution spectroscopy to measure Lithium
6708$\textrm{\AA}$ absorption equivalent widths.

The other alternative explanation for the X-ray active objects are
active binaries (ABs) of either RS CVn or BY Dra \citep{bopp77} type,
or cataclysmic variables. The studies of \citet{dempsey93} and
\citet{dempsey97} showed that $\log(L_X)$ can range up to 32 for RS
CVn and BY Dra systems, with $\log(F_X/F_V)$ up to -1.0. CVs also
typically have $\log(F_X/F_V)$ $\sim-$3-- +1 \citep{verbunt97},
however the lack of broad H$\alpha$ argues against CVs unless they
have sub-giant companions. To estimate the likely contribution of ABs
to our sample, we use a method similar to that used by
\citet{grindlay05}. We define a maximum distance at which an AB would
have been identified given the optical and X-ray detection limits of
our survey. Given some model for the distribution of ABs in the
Galaxy, we can predict the number we expect to detect in the
corresponding volume.

The detectability of a given AB in our {\it Chandra} observations is
determined by the detection limit of the observation, the hydrogen
column intervening between the telescope and the object, its
luminosity and the spectral model assumed for its emission.

To quantify the {\it Chandra} sensitivity in the calculation, we
extract the count rate limit for each observation across the field of
view. For simplicity we use an average value across the detector for
each field in the B$_X$ band. We derive the hydrogen column density
$N_H$ encountered as a function of distance using the D01 calculation
of $A_V$ in the direction of the aimpoint of each field. As a simple
approximation, we assume that ABs are distributed in the Galaxy with
some exponential scale height h in the z direction: n$_{CV}\propto
\exp^{-d(\sin |b|)/h}$, with n$_{AB}$ the AB space density and $b$ the
galactic latitude. This is probably a reasonable assumption for the
regions surveyed in this paper (provided we are only considering the
distribution within $\sim$3 kpc of the Solar neighborhood). Following
\citet{grindlay05}, we consequently utilize the formalism of
\citet{tinney93} in constructing an effective detection volume
V$_{eff}$, as defined by d$_{max}$:
\begin{equation}
V_{eff} = \Omega \left( h/ \sin |b| \right) ^3 \left[ 2 - \left( \chi
^2 + 2\chi + 2 \right) \exp \left( -\chi \right)\right]
\end{equation}
where $\chi$ = $d_{max} (\sin |b|)/h$ and $\Omega$ the solid angle
subtended by the ACIS field of view. V$_{eff}$ corrects the
$\emph{geometric}$ volume in which we search for the non-uniformity of
the AB space density. d$_{max}$ is the limiting distance at which a AB
at a luminosity of 10$^{32.0}$ ergs s$^{-1}$ \citep[the maximum value
for ABs found by ROSAT][]{dempsey93} could be detected in the {\it
  Chandra} observation considered. For the ACIS-S observation J1655,
we use the full 8$' \times$ 8$'$ field of view to calculate
$\Omega$. The GalCA pointing overlaps the SgrA$\star$ field of view,
so this field only adds three-quarters of the full ACIS-I solid angle
to the area of sky surveyed. The number of ABs we might expect to be
present in such a volume is then N$_{AB}$ = n$_{AB} \times$V$_{eff}$.

Following the conclusions of \citet{sciortino95} we adopt an AB scale
height of h = 250pc, and a local space density n$_{AB}$ =
3.7$\times$10$^{-5}$ pc$^{-3}$ \citep{favata95}. For the X-ray
spectral model, we adopt a MEKAL kT = 1.0 keV single temperature
model. To model the detection rate of ABs, we populate each volume
uniformly with a randomly distributed sample of 10$^4$ ABs from 0 pc
up to d$_{max}$ as determined for each field. Each is assigned an
X-ray luminosity and absolute visual magnitude, sampling randomly from
the data of \citet{dempsey93} and \citet{dempsey97}, using {\it only}
objects they classify as RS CVn or BY Dra type. This enables us to
determine if an object is luminous enough in X-rays and apparent
visual magnitude to be detected in our survey (using an assumed
detection limit of $V$=21 for an object to be spectrally
classified). Combining the detection rate derived with the predicted
number of ABs in each volume, we predict between 13 and 14 ABs found
in our survey. Based on absolute visual magnitude and high
$\log(F_X/F_{opt})$ ratio\footnote{RS CVn: $M_V$ more than 2$\sigma$
  less than that expected for a main sequence star of that spectral
  type and $\log(F_X/F_R)>-$5.0. BY Dra: $M_V$ consistent with main
  sequence, $\log(F_X/F_R)$ more than 2$\sigma$ greater than that
  expected for its spectral type \citep{huensch98ms}}, we flag our
best AB candidates in Table 5.

Our stellar coronal source sample appears likely to be a mix of both
local, young-MS and pre-MS stars, and a component of coronally active
binaries: RS CVn and BY Dra type. To specifically compare the stars of
these types in our sample with those found by ROSAT would require
significantly improved spectral, variability and orbital analysis to
more precisely classify our objects and tease out the contributions of
age, metallicity, and binarity that might also be contributing to the
observed differences the luminosity and X-ray to optical flux ratio.

\subsection{Constraints on the Galactic CV Density}
A main aim of the ChaMPlane survey is to investigate what constraints
we can place on the local CV space density. We follow a method similar
to that described in $\S$ 5.1.

For the {\it Chandra} sensitivity in the calculation, we use a single
value for the {\it Chandra} H$_C$ band, averaging over a 5$'$ radius
circle centered on the aimpoint. For the X-ray spectral model, we
adopt a kT = 8 keV bremsstrahlung emission spectrum as ``typical'' for
dwarf nova CVs \citep{warner95}. In this case, d$_{max}$ is the
limiting distance at which a CV at a luminosity of 10$^{32.5}$ ergs
s$^{-1}$ \citep[the maximum value for CVs found by ROSAT][]{verbunt97}
could be detected in the {\it Chandra} observation considered.

We adopt a scale height of h = 200pc, and following the conclusions of
\citet{grindlay05} among others, we adopt a local space density
n$_{CV}$ = 1$\times$10$^{-5}$ pc$^{-3}$. We populate each volume
uniformly with a randomly distributed sample of 10$^5$ CVs from 0 pc
up to d$_{max}$ as determined for each field. We assign each fake CV
an X-ray luminosity (in the H$_C$ band) and X-ray to optical flux
ratio $\log(F_x/F_V)$: we sample $L_x$ X-ray data as collected by
\citet{grindlay05}, originally presented in \citet{hertz90}, and the
ROSAT survey \citep[see:][]{verbunt97,schwope02}, to construct
distributions in $L_x$ and $\log(F_x/F_V)$ from which we randomly
sample. We assume that these two parameters are uncorrelated (a simple
scatter plot shows this to be the case for the 49 CVs in the ROSAT
sample). For a bremsstrahlung X-ray spectral model at 8keV, the
luminosity in the ROSAT band 0.1--2.4keV and our H$_C$ band is
approximately the same. We then derive an apparent $V$ magnitude and
observed X-ray flux. We set our photometric detection limit at a $V$
magnitude of 23---the faintest that we could have detected a CV via
its $H\alpha - R$ color. The number of CVs {\it expected} to be
present in each field is given in Table 8 as `CV$_{32.5}$' (column
5). The X-ray detection and optical identification percentages of
these objects are given in columns 6 and 7. The resultant number of
predicted CV detections is given as ID$_{32.5}$. Since the optical
detection limit is reached at only $\sim$2 kpc, the optical percentage
quoted below is essentially equal to the combined `X-ray-and-optical'
detection percentage.

\begin{deluxetable}{clccccccc}
\tablewidth{0pt} \tablecolumns{10} \tabletypesize{\footnotesize}
\tablenum{8} \tablecaption{CV Detection Constraints for the Five
Chandra ObsIDs} \tablehead{ \colhead{ObsID} & \colhead{Field} &
\colhead{CR limit\tablenotemark{a}} &
\colhead{dmax$_{32.5}$\tablenotemark{b}} &
\colhead{CV$_{32.5}$\tablenotemark{c}} &
\colhead{X-detect\tablenotemark{d}} &
\colhead{XO-detect\tablenotemark{e}} &
\colhead{ID$_{32.5}$\tablenotemark{f}} & \colhead{Found}\\ \colhead{}
& \colhead{} & \colhead{(ksec$^{-1}$)} & \colhead{(kpc)} & \colhead{}
& \colhead{(\%)} & \colhead{(\%)} & \colhead{} &\colhead{}} \startdata
99 & J1655 & 0.258 & 29.3 & 21.2 & 12.3$\pm$0.2 & 2.28$\pm$0.03 & 0.48 & 0(1)\tablenotemark{g}\\
737 & G347b & 0.305 & 15.75 & 37.3 & 4.1$\pm$0.2 & 0.65$\pm$0.03 & 0.24 & 0\\ 
944 & SgrB2 & 0.126 & 14.4 & 65.8 & 3.4$\pm$0.2 & 0.05$\pm$0.01 & 0.03 & 0\\ 
945 & GalCA & 0.305 & 11.3 & 22.7 & 3.0$\pm$0.2 & 0.13$\pm$0.02 & 0.03 & 2\\ 
53392 & SgrA$\star$ & 0.061 & 39.5 & 1232.0 & 1.7$\pm$0.02 & 0.014$\pm$0.002 & 0.17 & 1(1)\tablenotemark{g}\\ 
\hline Total & \nodata & \nodata &\nodata & 1378.8 & \nodata &\nodata & 0.95 & 3(5) 
\enddata
\tablecomments{\scriptsize{$^{a}$The average count rate limit for this
observation within 5$'$ of the $\it{Chandra}$ aimpoint. $^{b}$The maximum
distance we can detect a CV at $\log$(L$_x$)=32.5. $^{c}$The number of
CVs in the effective volume defined by the Tinney formula for this sky
position and d$_{max}$. $^{d}$The expected percentage of these CVs
detected in X-rays. $^{e}$The expected percentage of these CVs also
detected optically. $^{f}$The resultant number of CVs we expect to
identify in this field. $^{g}$This field includes one
spectroscopically unconfirmed CV, see Table 9.}}
\end{deluxetable}

It is apparent that there is an excess of CV candidates over the
number predicted by our simulation. In SgrA$\star$ and GalCA fields
considered in isolation there are a factor $\sim$10--70 too
many. However, summing over all fields we find between 3 and 5
detected, with 0.95$\pm$0.08 predicted. A simple $\chi ^2$ table shows
that this result is statistically significant at $\sim$92\%
confidence. What factors are contributing to the discrepant estimation
of the CV detection rate?

We have assumed that a single relationship between $A_V$ and distance
is applicable over each 16$' \times$ 16$'$ $\it{Chandra}$ field of
view yet extremes in the level of extinction are observed directly in
infrared images of the SgrA$\star$ and GalCA fields \citep[see
e.g.][]{laycock05}. The D01 model overlooks this small scale
variability across each field. If there were some covering factor of
higher-column-density gas and dust across each field or regions of
significantly lower extinction, we would expect to alter the number of
predicted CVs detected. However, a factor of $\sim$15 decrease in the
amount of extinction as a function of distance is necessary to produce
the factor $\sim$60 increase in the overall prediction for CV
detections in GalCA, and at least a factor $\sim$5 reduction is
required to solve the discrepancy for SgrA$\star$. Although there is
some great uncertainty in the dust model towards the Galactic Center
this appears unlikely. Improvements to the dust model are vital to
understand if this effect is more significant.

We included no radial component in our model CV space distribution yet
it is likely that there is some increase in the space density of CVs
as we approach the Galactic Center \citep[see, e.g.][]{rogel08}. Thus
we are underestimating the true number of CVs present in each volume,
before we apply our detection criteria. This may only be a minor
correction, since our optical detection and spectroscopic
identification limit restricts us to looking in the nearest $\sim$2
kpc to the Sun. Some additional work on improving how we model where
the CVs are in our volume, and how many we expect to be in this volume
is important to establish by how much we are underestimating
N$_{CV}$. Such a modification would appear to be most necessary for
the two fields closest to the Galactic Center: SgrA$\star$ and
GalCA. 

We restricted the detection solid angle to the inner 5$'$ of the
$\it{Chandra}$ field of view. Including the outer parts of the
detector (or for ACIS-S including the other S-chips) increases the
detection area by a factor $\sim$3. However, the X-ray count rate
limit beyond 5$'$ is lower (on average by a factor $\sim$2) than at
the aimpoint, which although not affecting optical detectability,
means the increase in predicted CV numbers would be small in
comparison with other effects discussed above.

Patterson (private communication) recommends a smaller scale height
h=150 pc on the basis of local CV surveys. Implementing this affects
only field J1655 significantly and would reduce the $\emph{predicted}$
number of CVs by $\sim$20\%---further in line with our lack of CVs
detected in this field.

At face value, our results suggest a higher in local value of the CV
density, n$_{CV}$, although alternatively the excess of objects
detected may simply represent fluctuations over a mean background
rate. The parameter n$_{CV}$ directly influences the predicted number
of CVs in any given field. Our sample is too small to be used to argue
strongly for a change, however the significance of the difference
between the observed number of CVs and our predicted number certainly
suggests an increase (by a factor $\sim$3--5) is justified. Two recent
studies find different values for n$_{CV}$. \citet{rogel08} construct
a Galaxy source-distribution model to predict CV detections found in
ChaMPlane Galactic Anti-center fields, and find a value of 10$^{-5}$
pc$^{-3}$ provides the best fit to their results. However,
\citet{ak08} use a sample of 459 local Solar-neighborhood CVs to
derive a value of n$_{CV}$ a factor $\sim$3 higher than this, which
appears to validate our own findings. Thus it may indeed be that the
value of n$_{CV}$ is somewhat higher than our adopted value.

\subsection{Resolving the Galactic Ridge X-ray Emission}
The extended Galactic Ridge X-ray Emission observed throughout the
Galactic Plane \citep[GRXE, see e.g.][]{worrall82} is the subject of
disagreement over whether deeper, higher resolution observations in
the X-ray band will eventually completely resolve all observed
Galactic emission at this wavelength, or whether in fact there is a
truly diffuse emitting plasma confined to the Galactic
Plane. \citet{ebisawa05} carried out deep, $\sim$100 ksec X-ray
observations of two fields in the Galactic Plane with {\it Chandra}
and estimated that point sources contribute at most 10\% of the total
X-ray flux they observe in this region. They conclude that no faint,
unobserved point source population could simultaneously match their
observed $\log$N--$\log$S (number versus source flux) plots of point
sources and the total GRXE flux, and hence there is a truly diffuse
component to the GRXE. On the other hand, \citet{revniv06} show that,
since the morphology of the GRXE very closely matches that of the
Galactic near-infrared surface brightness, it must trace the stellar
mass distribution of the Galaxy. They calculate the X-ray luminosity
per unit stellar mass that this conclusion requires in order to match
the observed flux of the GRXE.  Together with the X-ray luminosity
function of \citet{sazon06} they show that the X-ray emissivity per
unit mass of the local Solar neighborhood X-ray source population
extended to the whole Galaxy can easily account for the GRXE.

The probable types of the counterparts to our X-ray source population
are active binaries ($\sim$10\% of the sample), CVs ($\sim$2\%), YSOs
(perhaps between 39 and 61\%), with the remainder coronally emitting
stars, either young main sequence or main sequence objects. Broadly
and qualitatively this result agrees with the make-up of the Solar
neighborhood population as determined by \citet{sazon06}. A more
detailed comparison with their luminosity function is limited by two
factors: 1) the fact that our survey is not complete in either
magnitude/source flux space, or volume space, and 2) the lack of
detailed source classification and hence a determination of the X-ray
emissivity per stellar mass of our sources. However, our results
suggest a significant contribution to X-ray emitting sources close to
the Galactic Plane comes from pre-main sequence and young main
sequence stars, which can have relatively high X-ray luminosities
($L_x \sim 10^{31}$ erg s$^{-1}$). 

In order to make progress to address the conflict over the origin of
the GRXE requires one or both of: deep X-ray observations
\citep[$\sim$1 Megasec][]{revniv06} or accurate classification (via
optical or infrared spectroscopy or X-ray spectral analysis) of the
faint, hard X-ray source population present in many {\it Chandra}
X-ray studies of the Galactic Plane \citep[e.g.][]{muno04}. Although
our Galactic center pointing approaches the first requirement, it
still falls somewhat short and is hindered by the large extinction
towards this field ($\log (N_H) \approx 23.0$ cm$^{-2}$. Since we use
optical spectroscopy to classify targets, we are necessarily limited
to relatively bright ($R \lesssim$21), nearby objects and as a result
preferentially observe soft X-ray emitting, more unabsorbed sources. A
forthcoming ChaMPlane paper (Hong et al 2008, in preparation) will
utilise very deep ($\sim$1 Megasec) {\it Chandra} observations of the
Galactic center together with our quantile analysis technique to
classify faint point sources to directly consider the origin of the
GRXE.

\section{Conclusion}
We have carried out optical and X-ray spectral analysis on a sample of
X-ray detected optical sources in the Galactic plane, using a
combination of optical spectral fitting and quantile X-ray analysis to
obtain the extinction $E(B-V)$ and hence $A_V$ and hydrogen column
density, $N_H$ towards each object. We combine these estimates with
the work of \citet{drimmel01} who present a three-dimensional dust
model of the Galaxy in order to derive $A_V$ as a function of distance
in any direction, and thus further derive a distance to each object.

We present the discovery of a population of stellar coronal emission
sources, detected by $\it{Chandra}$ in five fields towards the
Galactic bulge. These are likely a mix of young stars, of roughly
Hyades and Pleiades age, as well as some pre-MS stars, and a component
of RS CVn or BY Dra type. We find no strong evidence that we have
sampled from stars with significantly different properties from local,
similarly active stars. We report the properties of the most probable
RS CVn and BY Dra-type candidates from our sample, and identify one
possible qLMXB candidate also. We note that this latter object could
instead be an RS CVn system. High resolution optical spectra can make
this clear.

We report the discovery of three X-ray detected CVs in the direction
of the Galactic Center. All three are consistent with having an X-ray
spectrum consisting of bremsstrahlung at kT $\approx$ 8keV, and are
within $\approx$2 kpc of the Sun. An additional 2 CVs are indicated by
our photometry and X-ray data, and can be tested with optical
spectroscopy.

The number of CVs detected in our survey is consistent with a local CV
space density of $\sim$10$^{-5}$ pc$^{-3}$, and a scale height
$\sim$200pc, but is suggestive of a larger local value or strong
radial gradient. However, there is considerable uncertainty in the
model we use to predict extinction as a function of distance and hence
derive the number detected in our survey. Although the numerical
uncertainty in the model appears to be a factor of $\sim$3 (see Figure
5), cf. the required factor $\sim$5--15 to rectify the discrepancies
seen in the GalCA and SgrA fields, it is possible that true variations
in the distribution of dust in the Galaxy might be able to explain
this. Further work on better modelling the Galactic dust distribution
and CV content of our fields is desirable to improve our constraints.

\acknowledgements The author would like to thank John Silverman for
collecting the LDSS2 spectral data during the 2001 run; JEG collected
the LDSS2 spectra in the 2002 run. We also thank the two anonymous
referees and Eric Feigelson whose suggestions considerably improved
the manuscript. This research has made use of the SIMBAD database,
operated at CDS, Strasbourg, France and the NASA Astrophysics Data
System. This work is supported in part by NASA/$\it{Chandra}$ grants
AR1-2001X, AR2-3002A, AR3-4002A, AR4-5003A, AR6-7010X, NSF grant
AST-0098683, and the $\it{Chandra}$ X-ray Center. We thank NOAO for
its support via the Long Term Survey program.

\clearpage

\begin{figure}
\begin{center}	
\includegraphics[width=5.6in]{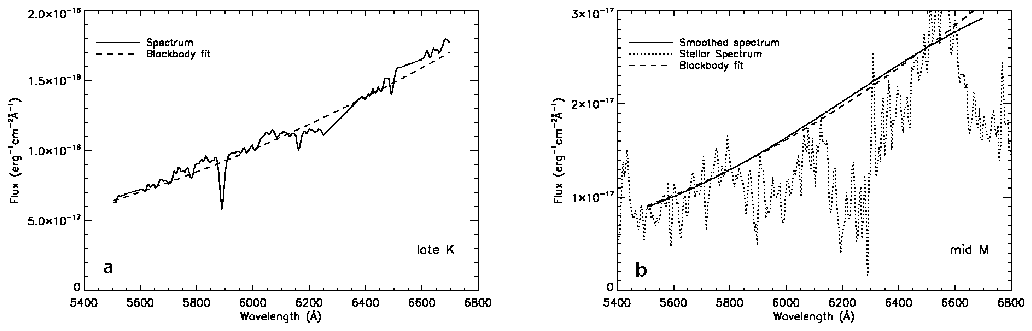}
\caption{Two example fits to spectra from the LDSS2 sample. The panel
  for the M star (right) also shows the polynomial fit to the spectrum
  (see text for description) to which the blackbody curve was
  subsequently fit.}
\end{center}
\end{figure}

\begin{figure}[ht]
\begin{center}
\includegraphics[width=6in]{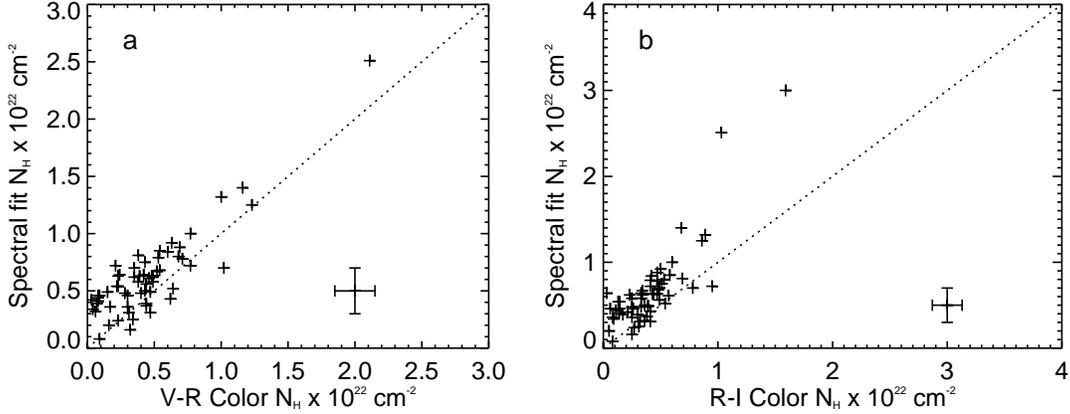}
\caption{Plots of hydrogen column density for all stars with available
  $V-R$ or $R-I$ photometry in the sample, derived from the spectral
  fit and color methods detailed in $\S$$\S$ 3.1 and 3.2. Typical
  error bars and lines of $N_H$(Color)=$N_H$(Fit) (dotted line) are
  shown for reference.}
\end{center}
\end{figure}

\begin{figure}[ht]
\begin{center}
\includegraphics[width=3.5in]{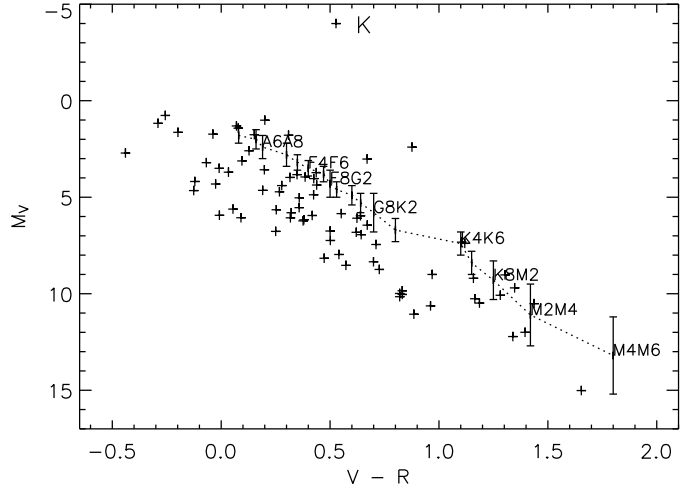}
\caption{A Color-Magnitude diagram of the stars in the LDSS2
sample. We plot M$_V$ = $V$ $-5\log$(dist) + 5 $- A_V$, and $(V-R)_0$
= $V-R - 0.781E(B-V)$. The star marked `K' is the K giant in SgrB2,
ChOPSJ174642.22$-$282907.4. The dotted curve and spectral type labels are from \citet{Allen2000} and the error bars show the range of M$_V$ for spectral type range plotted.}
\end{center}
\end{figure}

\begin{figure}
\begin{center}
\includegraphics[width=6.5in]{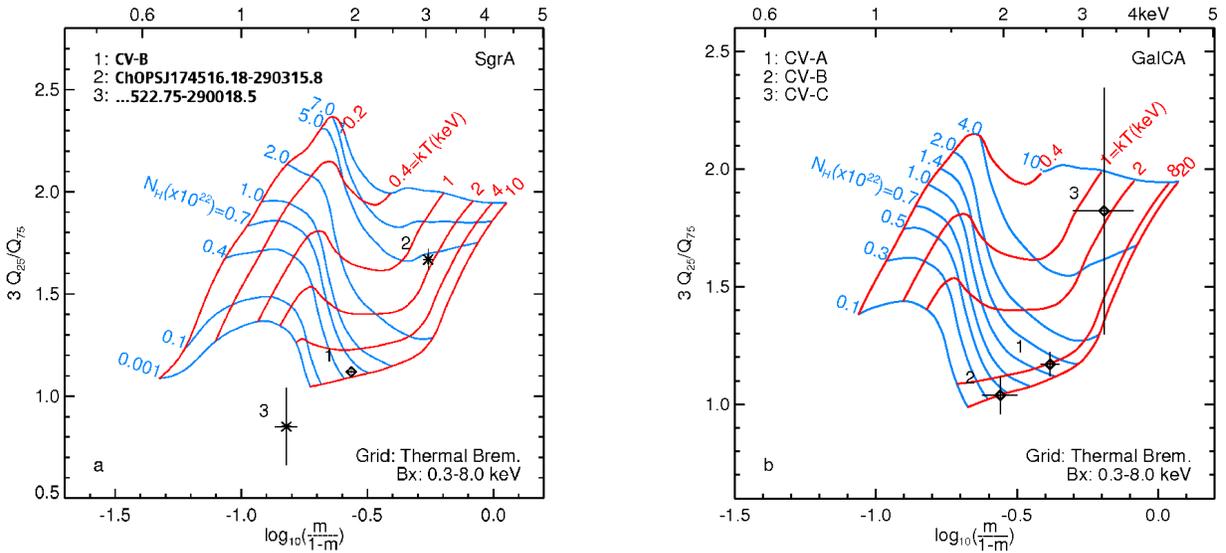}
\caption{Left panel: QCCD plot for ObsID 53392 (SgrA$\star$) showing
  the positions of CV-B and non-identified sources from the
  SgrA$\star$ field with at least 107 counts. Right panel: ObsID 945
  (GalCA) showing all three CVs.}
\end{center}
\end{figure}

\begin{figure}[ht]
\begin{center}	
\includegraphics[width=6in]{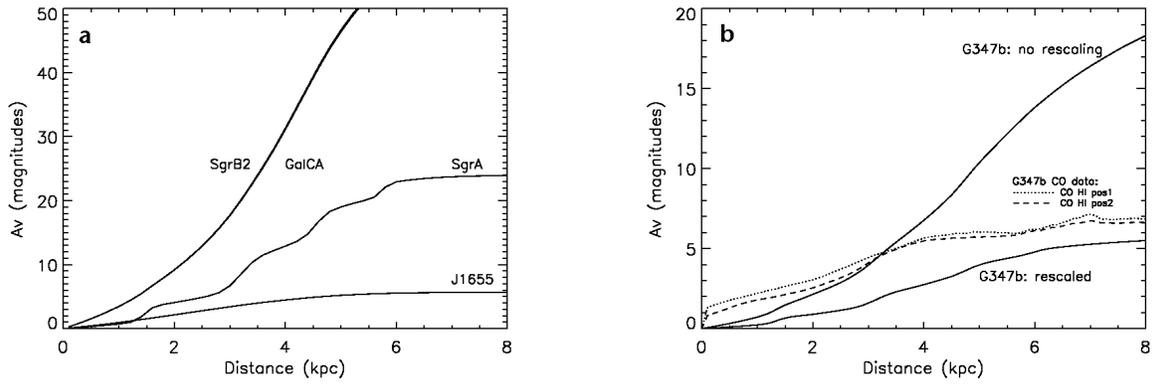}
\caption{\footnotesize{Plots of $A_V$ (re-scaled values) versus
distance from the model of \citet{drimmel01} for the five
$\it{Chandra}$ fields. For field G347b (right-hand plot) we show the
results of both rescaled and non-rescaled $A_V$ versus distance from
this paper, and also overplot our own results, as derived from CO+HI
observations. See $\S$ 3.2 for a description.}}
\end{center}
\end{figure}

\begin{figure}
\begin{center}
\includegraphics[width=3.5in]{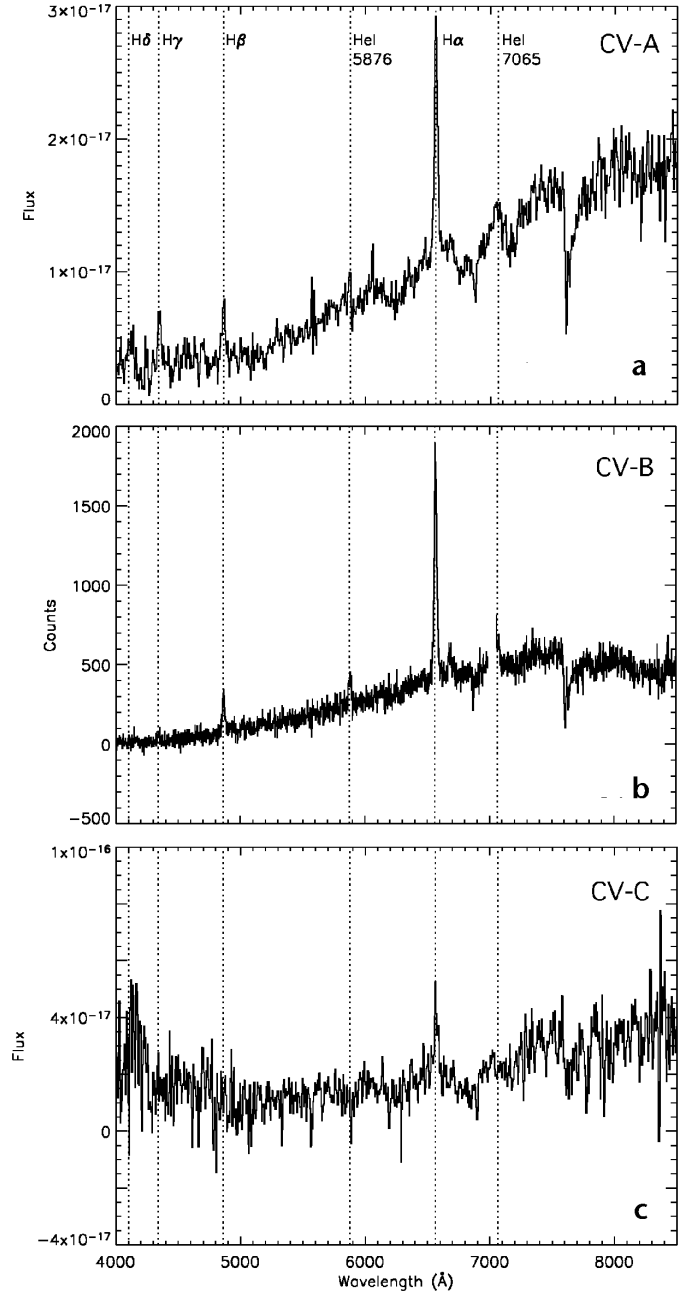}
\caption{The three CVs discovered in the five $\it{Chandra}$ fields in
this paper. The LDSS2 spectra for CV-A and CV-C have y-axis
units ergs s$^{-1}$ cm$^{-2}$ $\textrm{\AA}$. The IMACS spectrum
for CV-B was not flux calibrated and has y-axis units in raw
counts. The region 6990--7500$\textrm{\AA}$ has been removed from CV-B
as it covers a CCD chip gap.}
\end{center}
\end{figure}

\begin{figure}
\begin{center}
\includegraphics{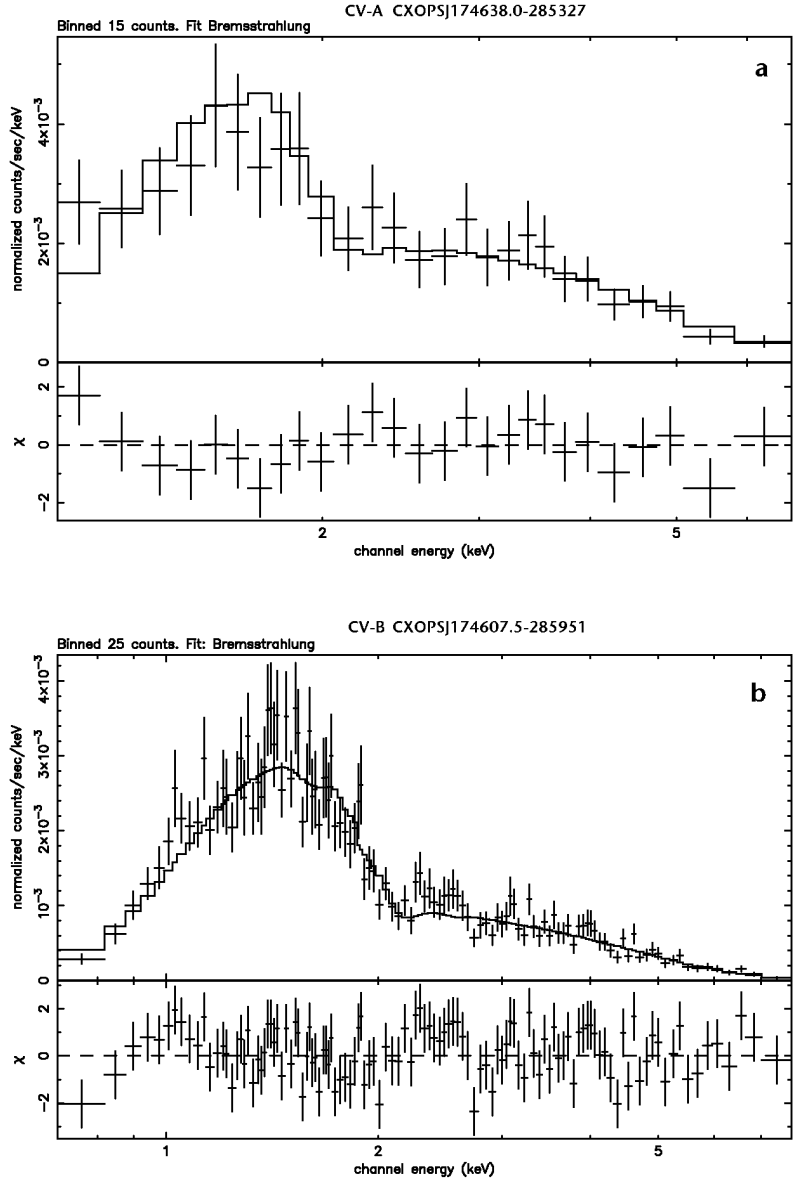}
\caption{XSPEC fits to the X-ray data for the CV-A and CV-B detected
in this survey. Upper spectrum, CV-A: 438$\pm$22 net counts (B$_X$ band),
lower spectrum, CV-B: 3539$\pm$63 net counts. We plot the spectrum plus
fit, and in the lower panel in each case the residuals of the spectrum divided by the errors.}
\end{center}
\end{figure}

\begin{figure}[ht]
\begin{center}	
\includegraphics[width=6.4in]{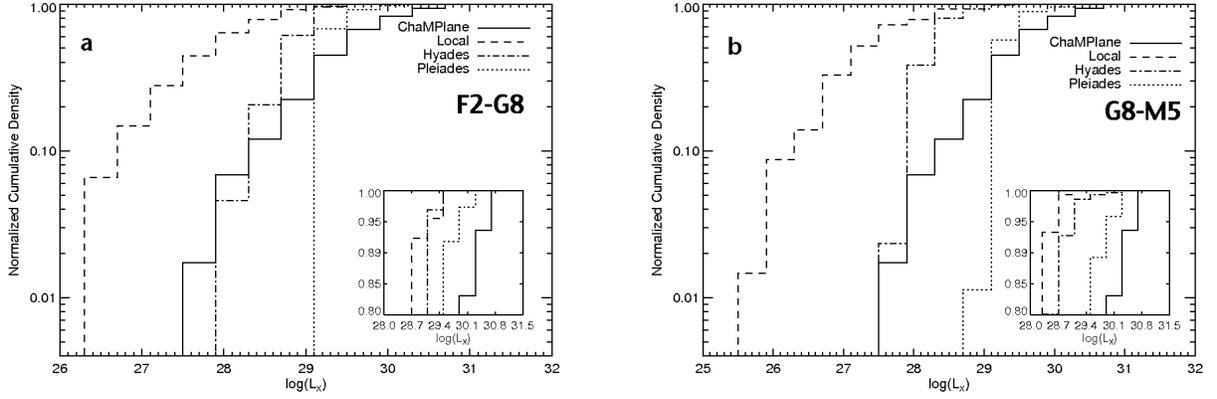}
\caption{X-ray luminosity functions of our stellar coronal sample,
  1/V$_{max}$ corrected. We divide our sample into stars from type
  F2--G8 (58 stars, left panel) and G8--M5 (48 stars, right
  panel). Inset in each Figure is a zoomed in portion of the top part
  of the plot. We overplot comparison stellar samples from the a local
  sample, plus the Hyades and Pleiades.}
\end{center}
\end{figure}

\begin{figure}
\begin{center}
\includegraphics[width=6.3in]{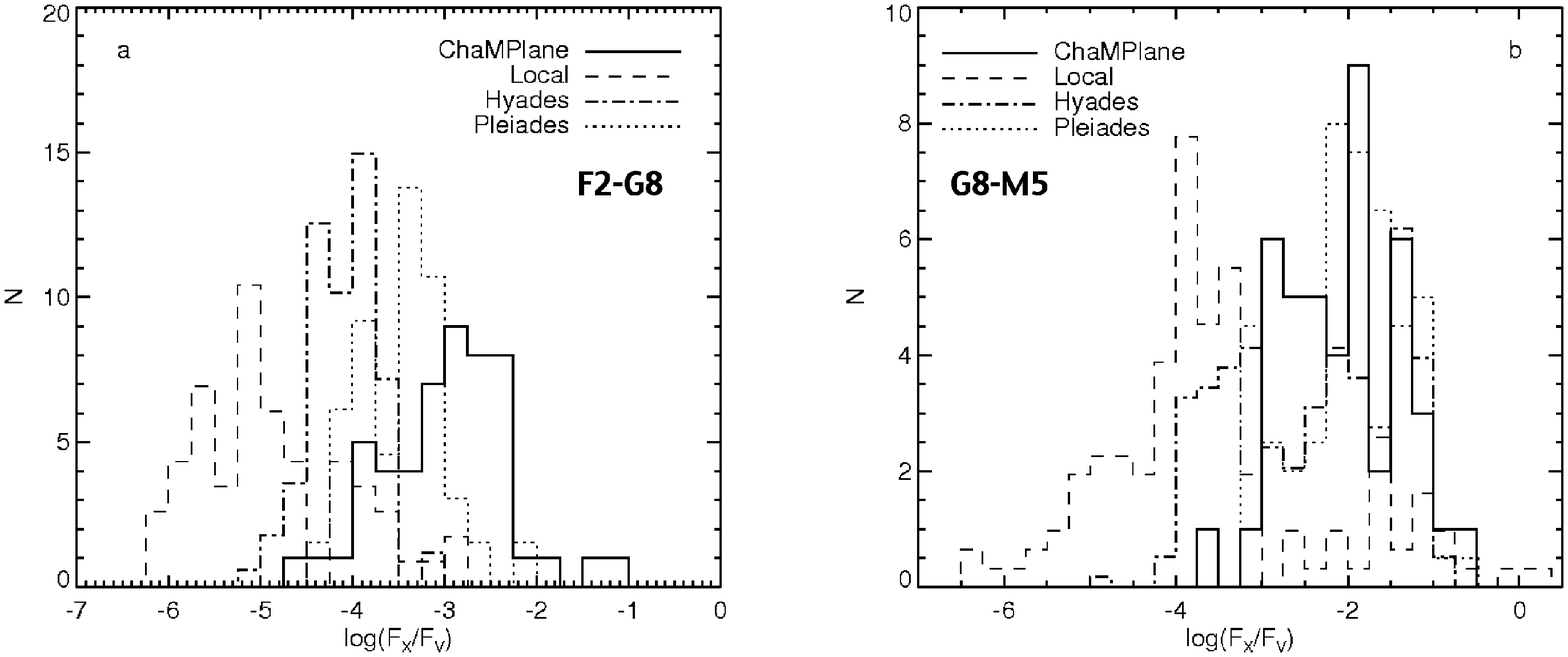}
\caption{Histograms by spectral type grouping (same as Figure 7) of
  the X-ray to optical flux ratio of our stellar sample, overplotted
  with data from the studies of \citet{schmitt04} (dashed) and the
  Hyades and Pleiades samples of \citet{stern95} and
  \citet{micela96}. We classify 62--77\% of our M stars as dMe from
  their H$\alpha$ emission.}
\end{center}
\end{figure}

\begin{figure}
\begin{center}
\includegraphics[width=5in]{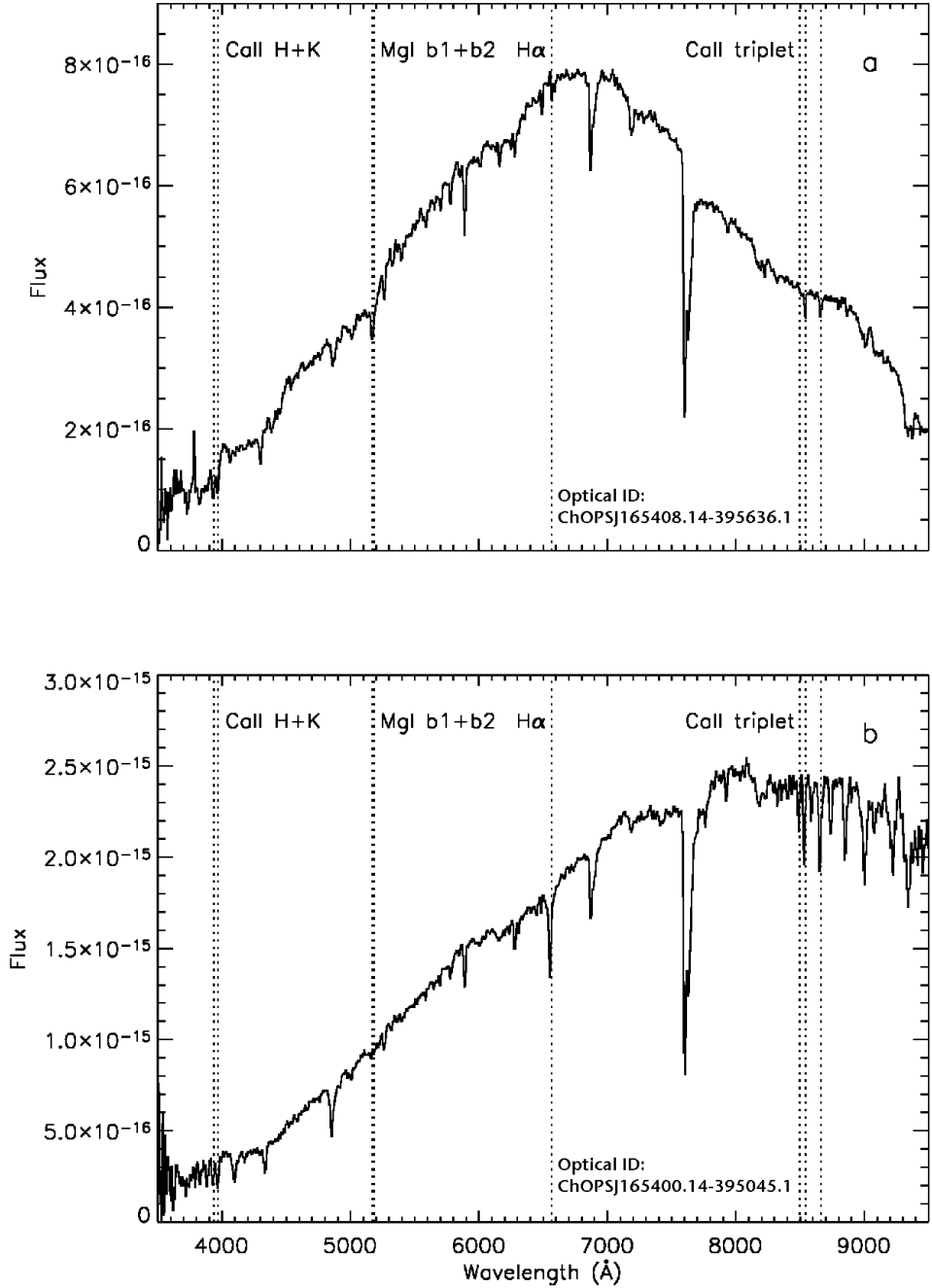}
\caption{Upper spectrum: the qLMXB candidate found in our LDSS2
sample. Lower spectrum: GRO J1655 $-$40 as observed by LDSS2 from our
June 2002 observing run. The flux scale has units: ergs s$^{-1}$
cm$^{-2}$ $\textrm{\AA} ^{-1}$. Important spectral lines are marked on
the spectra for reference. The strongest spectral feature at $\lambda$7600 
is telluric absorption by the Earth's atmosphere.}
\end{center}
\end{figure}

\begin{figure}
\begin{center}
\includegraphics[width=5in]{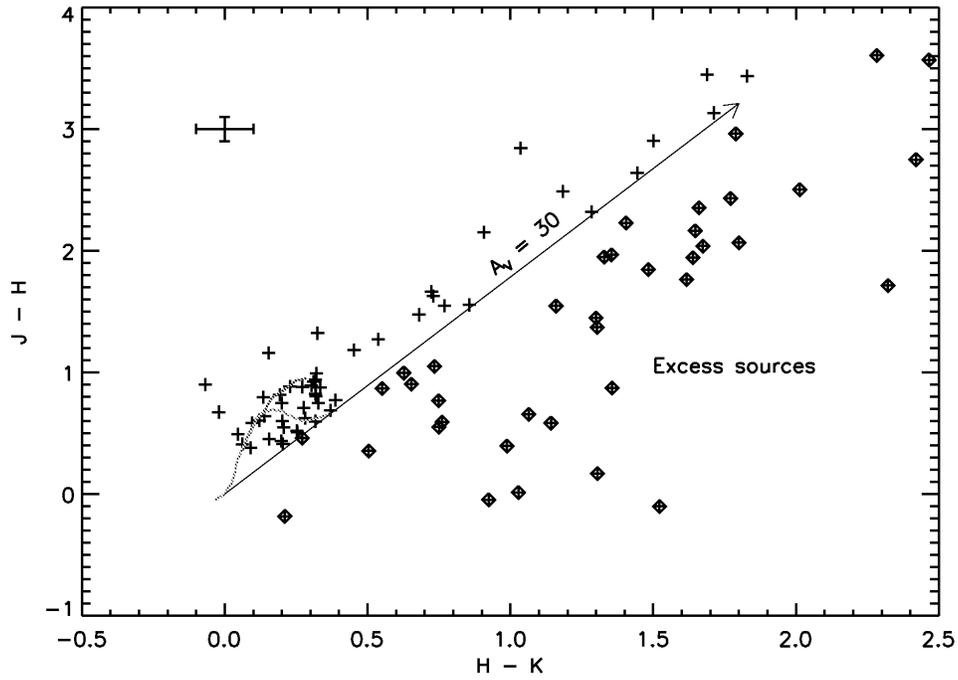}
\caption{$J-H$ versus $H-K_S$ plot of 2MASS sources matched with
  ChaMPlane sources in this paper. Overlaid is the locus of dwarf and
  giant stars (giant stars are redder) from \citet{bessell88}. The
  arrow represents an extinction of $A_V$=30 using the extinction
  relation of \citet{nishi06}. Sources marked with diamond points are
  candidate excess sources/young stars. A typical error bar is shown,
  upper left.}
\end{center}
\end{figure}

\end{document}